# Computational screening of repurposed drugs and natural products against SARS-Cov-2 main protease ($M^{pro}$) as potential COVID-19 therapies

Sakshi Piplani[1-2*], Puneet Singh[2*], Nikolai Petrovsky[1-2#] David A. Winkler[3-6#]

1 College of Medicine and Public Health, Flinders University, Bedford Park 5046, Australia

2 Vaxine Pty Ltd, 11 Walkley Avenue, Warradale 5046, Australia

3 La Trobe University, Kingsbury Drive, Bundoora 3042, Australia

4 Monash Institute of Pharmaceutical Sciences, Monash University, Parkville 3052, Australia

5 School of Pharmacy, University of Nottingham, Nottingham NG7 2RD. UK

6 CSIRO Data61, Pullenvale 4069, Australia

[*] contributed equally as first authors

# contributed equally as last authors

## Abstract

There remains an urgent need to identify existing drugs that might be suitable for treating patients suffering from COVID-19 infection. Drugs rarely act at a single molecular target, with off target effects often being responsible for undesirable side effects and sometimes, beneficial synergy between targets for a specific illness. Off target activities have also led to blockbuster drugs in some cases, e.g. Viagra for erectile dysfunction and Minoxidil for male pattern hair loss. Drugs already in use or in clinical trials plus approved natural products constitute a rich resource for discovery of therapeutic agents that can be repurposed for existing and new conditions, based on the rationale that they have already been assessed for safety in man. A key question then is how to rapidly and efficiently screen such compounds for activity against new pandemic pathogens such as COVID-19. Here we show how a fast and robust computational process can be used to screen large libraries of drugs and natural compounds to identify those that may inhibit the main protease of SARS-Cov-2 (3CL pro, $M^{pro}$). We show how the resulting shortlist of candidates with strongest binding affinities is highly enriched in compounds that have been independently identified as potential antivirals against COVID-19. The top candidates also include a substantial number of drugs and natural products not previously identified as having potential COVID-19 activity, thereby providing additional targets for experimental validation. This *in silico* screening pipeline may also be useful for repurposing of existing drugs and discovery of new drug candidates against other medically important pathogens and for use in future pandemics.

## Introduction

The devastating impact of the COVID-19 pandemic caused by SARS coronavirus-2 (SARS-CoV-2) has stimulated unprecedented international activity to discover effective vaccines and drugs for this and other pathogenic coronaviruses such as SARS and MERS CoV.[1-17] Computational methods offer considerable promise for determining the affinities of small drug-like molecules for SARS-Cov-2 protein targets. Recent papers in Science have reported the application of computational de novo drug design based on the structures of the SARS-Cov-2 protease. [18, 19] Clearly, design of potent new drugs for coronaviruses is very important for future pandemic preparedness, given that the last three serious epidemics have been caused by coronaviruses. However, to make an impact on the current COVID-19 pandemic, it is only feasible to use drugs that are already registered (off label use), have been through at least phase 1 clinical trials to establish initial human safety, or are approved natural products. Any COVID-19 drug candidates identified in this way can then be used very quickly, as their safety and pharmacokinetics should be already well understood. Drugs that reduce viral replication primarily by targeting viral proteases and polymerases are classified as direct-acting antivirals and are the focus of the current work. Other studies have explored host-targeted drugs that inhibit cellular functions required for viral replication and thereby inhibit SARS-Cov-2 infection, albeit with more potential for host side effects.[20]

The SARS-Cov-2 genome encodes > 20 proteins, many of which are potential antiviral drug targets (Figure 1). Two proteases (PL pro and 3CL pro) are essential for virus replication. These enzymes cleave the PP1A and PP1AB polyproteins into functional components. 3-chymotrypsin-like protease (3CLpro, aka main protease, $M^{pro}$) catalytically self-cleaves a peptide bond between a glutamine at position P1 and a small amino acid (serine, alanine, or glycine) at position P1'. This protease corresponds to non-structural protein 5 (nsp5), the main protease in coronaviruses. 3CL protease is crucial to the processing of the coronavirus replicase polyprotein (P0C6U8), cleaving it at 11 conserved sites. 3CL protease employs a cys-his catalytic dyad in its active site where the cysteine sulfur is the nucleophile and the histidine imidazole ring is a general base. Very recent research has shown that strong $M^{pro}$ inhibitors can substantially reduce SARS-Cov-2 virus titres, reduce weight loss and improve survival in mice,[21] making $M^{pro}$ a promising drug target for structure-based drug discovery.

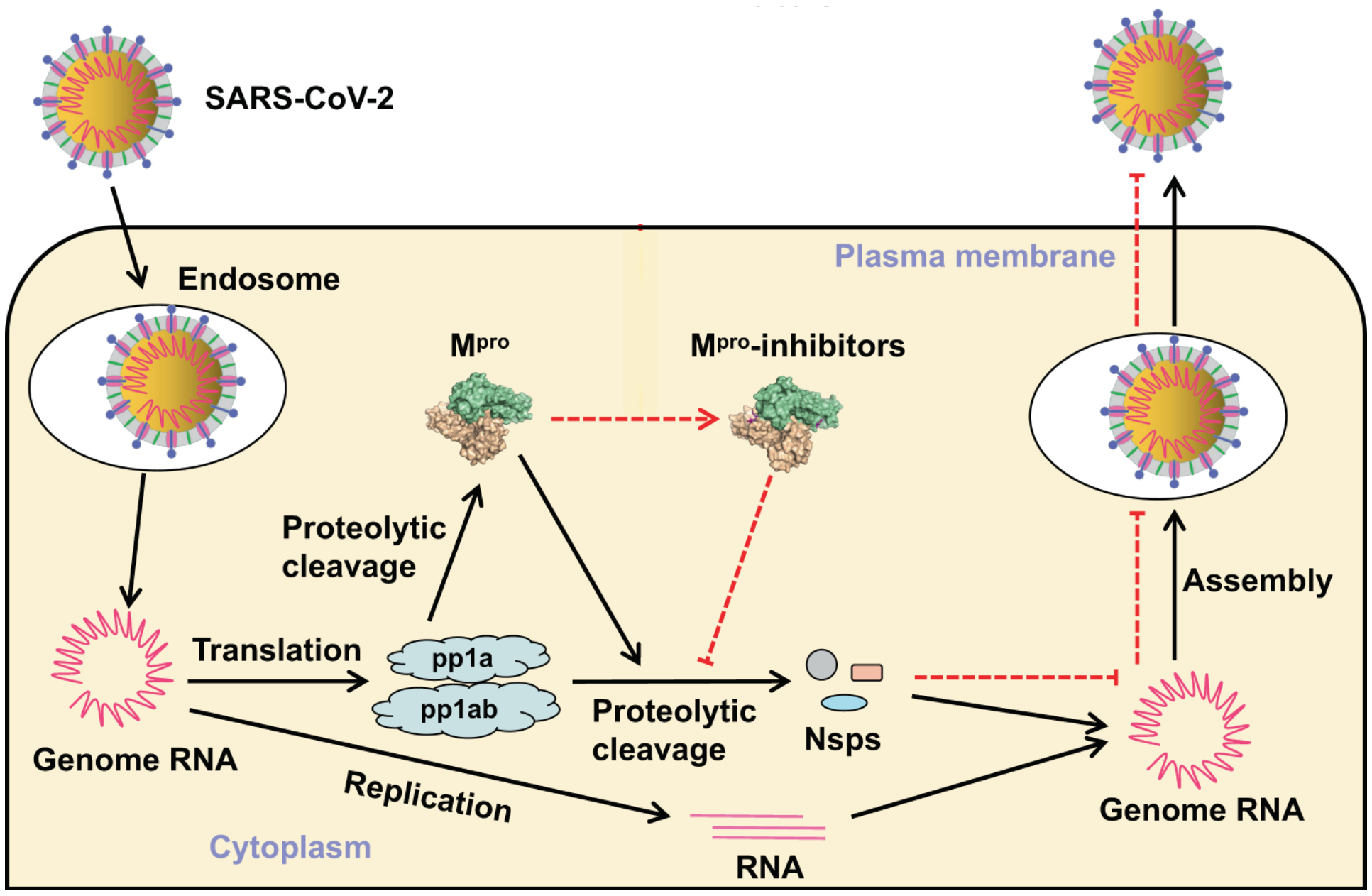

**Figure 1**. Virus entry and replicative cycle. $M^{pro}$ produces non-structural proteins, Nsps, that are essential for assembly of the viral replication transcription complex needed for RNA synthesis. Inhibitors bind to $M^{pro}$, resulting in failure of virion assembly and inhibition of release of new functional virions. Adapted from Mengist et al.[22] https://creativecommons.org/licenses/by/4.0/

Computational methods can rapidly and efficiently identify candidate drugs for repurposing in pandemic situations where speed is of utmost importance. Here we used molecular docking followed by high throughput molecular dynamics simulations to prioritize from an initial large number of licensed or clinical trial drugs and natural products, a short list of the most promising candidates. Molecular dynamics calculations were used to predict the optimal binding poses and binding energies for 84 of the top hits from docking-based virtual screening to the SARS-CoV-2 $M^{pro}$. Finally, the top candidates were ranked based on binding affinity and novelty, for COVID-19 repurposing.

## Results and discussion

The binding energies of the 84 top ranked ligands from the docking calculations are listed in Supplementary Table 1. The ten drugs with the tightest binding to $M^{pro}$ are summarized in Table 1, together with their GMXPBSA binding energies. The calculated binding energy of several of the antiviral drugs, namely, simeprevir, sofosbuvir, lopinavir, and ritonavir are very similar, within the uncertainties in calculated energies. Several of the top hits were antivirals identified in other *in silico* docking studies or wet-lab SARS-CoV-2 activity studies. This provides a degree of validation that our computational methods are appropriate and are yielding similar results to other published studies for some well-studied antiviral drugs. The web site DrugVirus.info provides a concise picture of the broad-spectrum antiviral activity of a range of drugs; a summary for four of the antiviral hits from our *in silico* screens is provided in Figure 2.

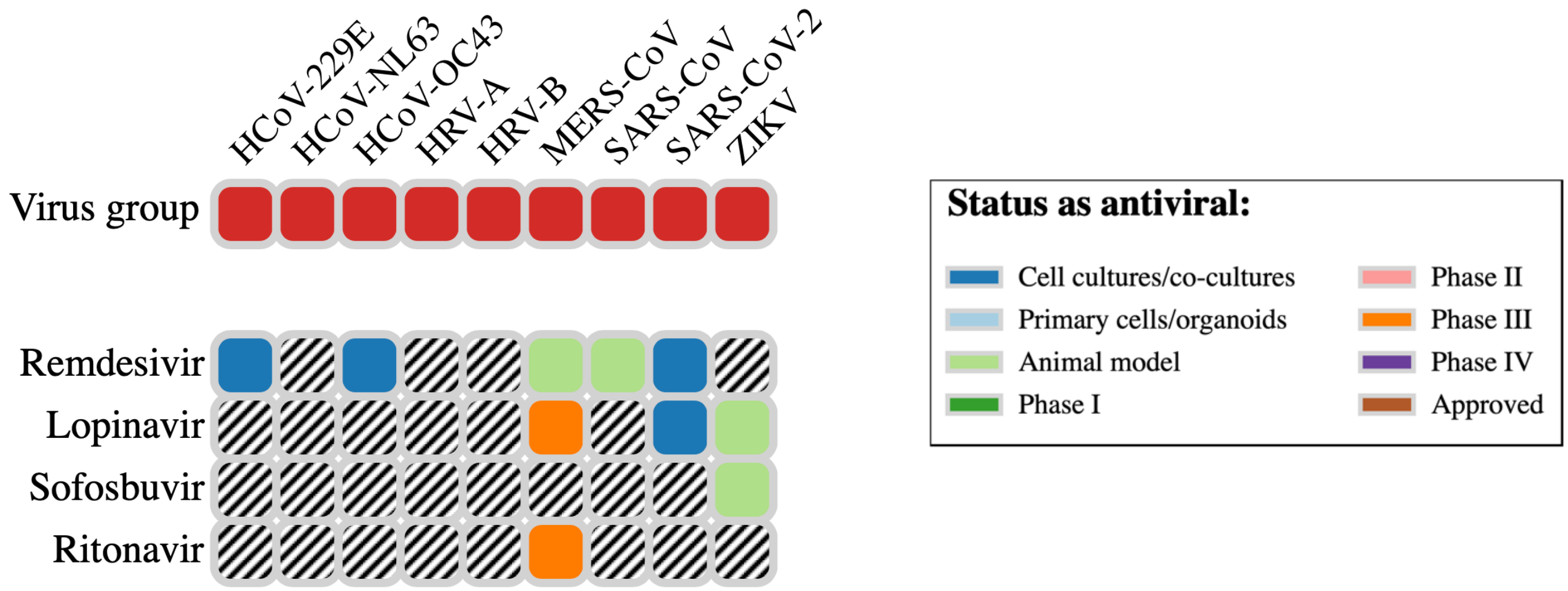

**Figure 2**. Spectrum of antiviral activity and nature of assessment for four antiviral hit drugs.

Simeprevir was reported to be an inhibitor of the 3CLPro protease by Abhithaj et al.[23] They used a pharmacophore search followed by grid-based ligand docking (GLIDE, Schrodinger) and binding energy estimates from the MMGBSA method of -81.7 kcal/mol. However, they did not use MD to simulate the interaction of Simeprevir in the 3CLPro binding site. Similarly, Sofosbuvir was reported to be a strong inhibitor of the protease by Lo et al.[24] Using a Vero E6 cellular infection model, they also reported that Simeprevir was the only drug among their prioritized candidates that suppressed SARS-CoV-2 replication at below 10 μM. Dose-response studies showed that Simeprevir had an $EC_{50}$ of 4 μM, and $CC_{50}$ of 20 μM, similar to Remdesivir in their experiments.

**Table 1**. Binding energies of 10 top ranked predicted small molecule ligands to SARS-Cov-2 $M^{pro}$.

| ID | Structure | Description | $\Delta G_{MMPBSA}$ ($\Delta G_{bind}$) (kcal/mol) |
|---|---|---|---|
| C3809489 Bemcentinib | 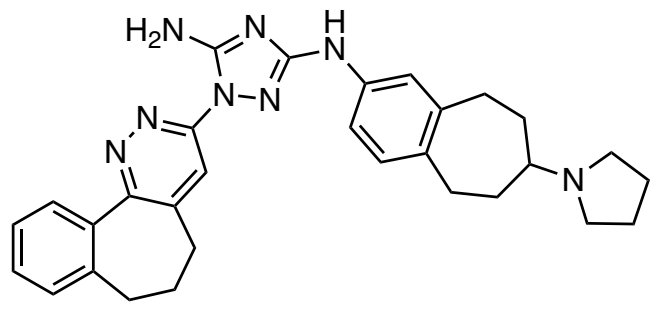 | Inhibitor of the kinase domain of AXL receptor. | -34.7±2.6 (-30.7) |
| C4291143 PC786 | 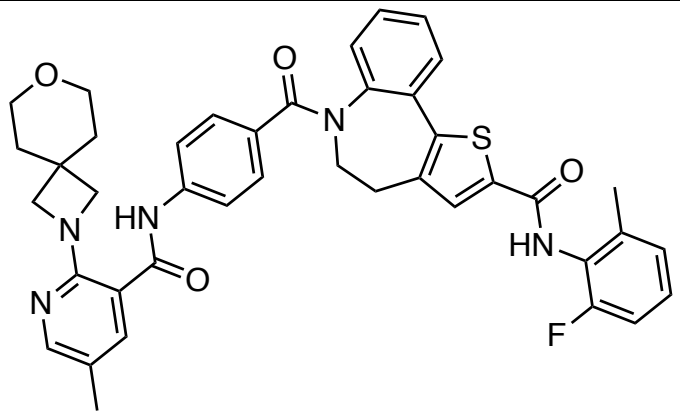 | Respiratory syncytial virus (RSV) L protein polymerase inhibitor. | -33.1±0.3 (-29.2) |

| ID | Structure | Description | $\Delta G_{MMPBSA}$ ($\Delta G_{bind}$) (kcal/mol) |
|---|---|---|---|
| C787 Montelukast | 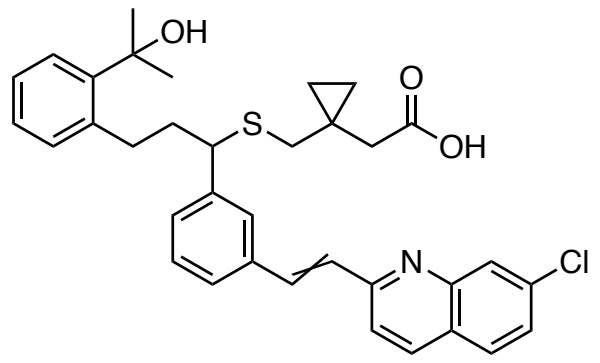 | Leukotriene receptor antagonist used with cortico-steroids for asthma therapy. | -32.7±0.2 (-20.6) |
| C442 Ergotamine | 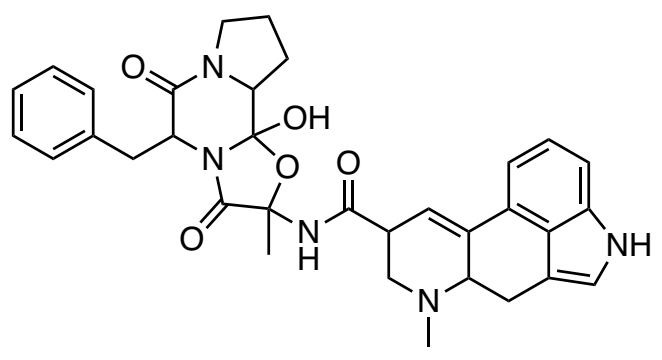 | Alpha-1 selective adrenergic agonist used in migraine treatment. | -31.5±0.3 (-28.7) |
| D06290 Simeprevir | 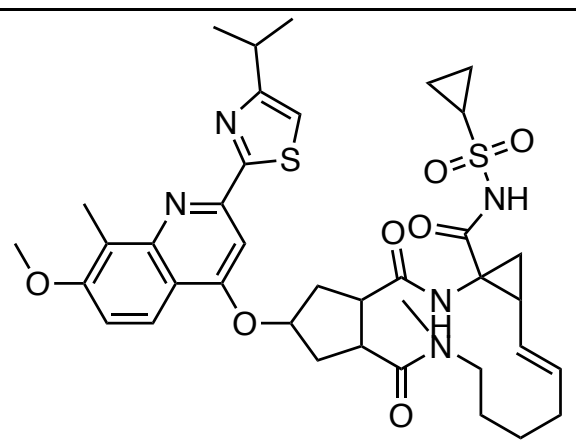 | Hepatitis C virus (HCV) NS3/4A protease inhibitor. | -31.4±0.2 (-29.2) |
| D08934 Sofosbuvir | 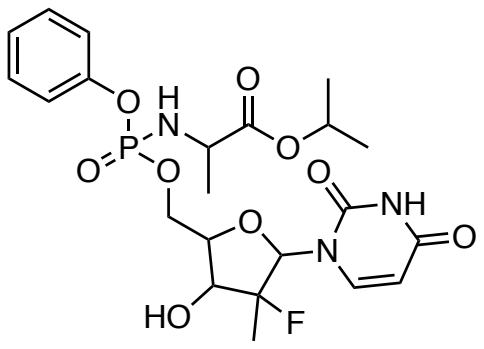 | Nucleotide prodrug and HCV NS5B polymerase inhibitor | -31.0±0.5 (-22.8) |
| D01601 Lopinavir | 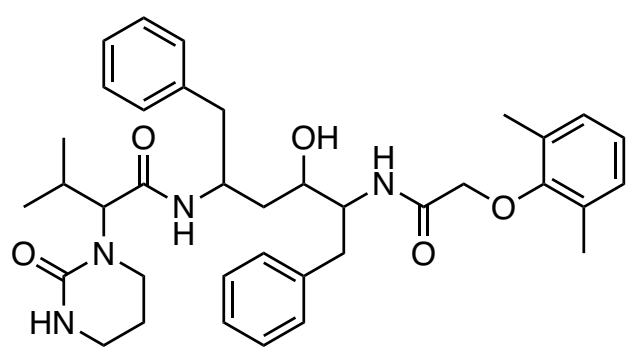 | Antiretroviral protease inhibitor for treatment of HIV-1 | -30.7±0.3 (-20.4) |
| D00503 Ritonavir | 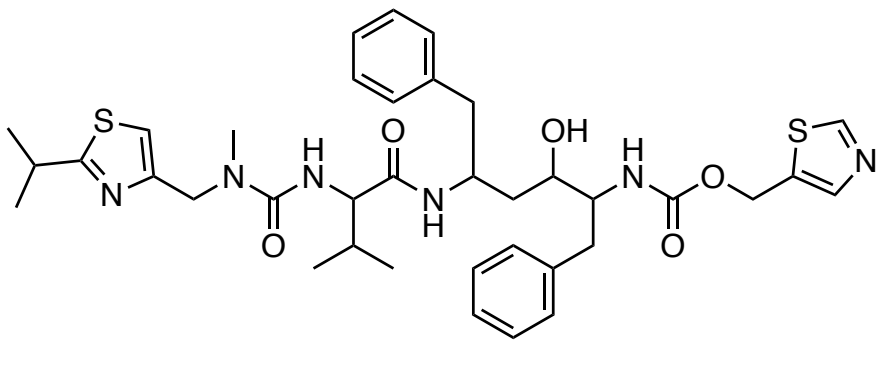 | Peptidomimetic inhibitor of HIV-1 and HIV-2 proteases | -30.5±0.5 (-21.3) |

| ID | Structure | Description | $\Delta G_{MMPBSA}$ ($\Delta G_{bind}$) (kcal/mol) |
|---|---|---|---|
| C2105887 Mergocriptine | 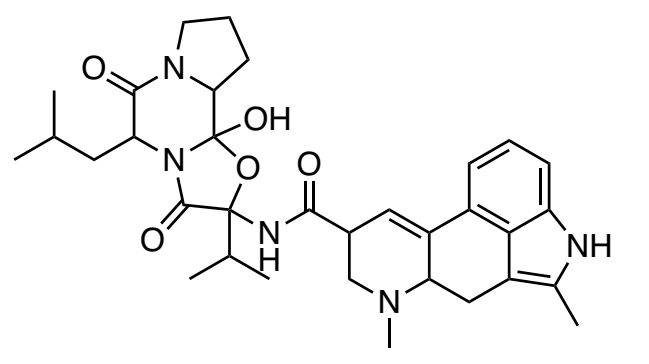 | Synthetic ergot derivative, dopamine receptor agonist. | -30.0±0.3 (-17.9) |
| D14761 Remdesivir | 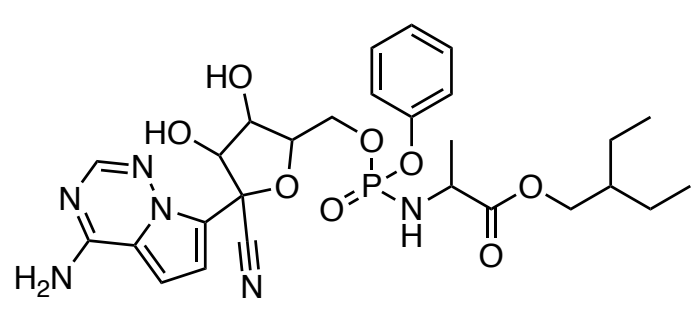 | Viral RNA-dependent RNA polymerase inhibitor. | -30.0±0.2 (-27.1) |

The potential protease inhibition properties of Lopinavir and Ritonavir were reported by Bolcato et al., who used supervised MD to calculate the trajectories of the ligands in the protease binding site.[25] Costanzo and colleagues likewise reported high protease binding for these two antiviral drugs.[26] They also reported updates on experimental drugs successfully employed in the treatment of the disease caused by SARS-CoV-2 coronavirus. Patient recovery has been reported after treatment with lopinavir/ritonavir (used to treat HIV infection) in combination with the anti-flu drug oseltamivir. Muralidharan et al. also used AutoDock (another docking program similar to Vina produced by the Scripps group) followed by MD simulations using the Generalised Amber Force Field (GAFF) in Amber16 to screen for repurposed drugs[27] They reported AutoDock binding energies for lopinavir, oseltamivir and ritonavir of −4.1 kcal/mol, −4.65 kcal/mol, −5.11 kcal/mol, respectively but did not provide the binding energies from the MD calculations. The best-known antiviral drug, which has been the subject of several clinical trials for COVID-19, is Remdesivir.[28] The potential inhibition of the protease by this drug has been reported by several computational screening studies. For example, Al-Khafaji and colleagues reported a combined computational docking and MD study of a range of antiviral drugs to the viral protease.[29] They calculated a binding energy for remdesivir of −65.19 kcal/mol from a GROMACS simulation and a MMGBSA binding energy calculation. Beck et al. reported a $K_d$ for binding of remdesivir to 3CLPro of 113 nM using a deep learning model. Liu et al reported an *in vitro* assay that exploited the pronounced cytopathic effects of SAR-Cov-2 on Vero cells and the ability of a range of antiviral drugs to protect cells against the virus.[30] In their assay, Remdesivir exhibited an $IC_{50}$ of 2.5μM and $CC_{50}$ of 175μM, while Sofosbuvir, Lopinavir and Ritonavir were inactive. Similarly, Ma et al. reported a fluorescence resonance energy transfer (FRET)- based enzymatic assay for the SARS-CoV-2 $M^{pro}$ and applied it to screening a library of protease inhibitors.[31] In their assay, Simeprivir exhibited an $IC_{50}$ of 14±3 μM.

The most interesting potential protease inhibitors from our study are the ergot alkaloids ergotamine, mergocriptine, the thrombopoietin receptor agonist eltrombopag (ranked 13 with $\Delta G_{MMPBSA}$=–28.2 kcal/mol, see Supplementary Table 1), bemcentinib, PC786, and montelukast. These drugs were predicted to have better binding energies than the antiviral drugs discussed above and have a higher degree of novelty.

*Ergot drugs*

Gurung et al. reported potential binding of ergotamine to the SAR-Cov-2 main protease in a preprint.[32] The employed AutoDock Vina but without subsequent MD simulation of the complex. They reported the binding energy as −9.4 kcal/mol for dihydroergotamine and -9.3 kcal/mol for ergotamine. Mevada et al. also reported the in-silico estimates of the binding of ergotamine to the protease using AutoDock Vina for the virtual screening.[33] They found the drug bound with an energy of -10.2 kcal/mol, calculated using Vina (no subsequent MD simulation). Gul et al. used a similar docking approach, this time with MD simulation, and identified ergotamine and its derivatives dihydroergotamine and bromocriptine to have high binding affinity to SARS-Cov-2 3CLpro. Ergotamine is an alpha-1 selective adrenergic agonist and vasoconstrictor that had a high docking binding energy against SARS-Cov-2 $M^{pro}$ of -8.6 kcal/mol. Dihydroergotamine, the 9,10-alpha-dihydro derivative of ergotamine, showed similar high affinity of -8.6 kcal/mol and bromocryptine -9.2 kcal/mol. Ergotamine has also been predicted to bind tightly to the SARS-Cov-2 spike (S) protein.[34] Figure 3 shows a LigPlot representation of the interactions of key functional groups in ergotamine and mergocriptine with protease active site residues. These are also listed in Supplementary Table 2 for reference.

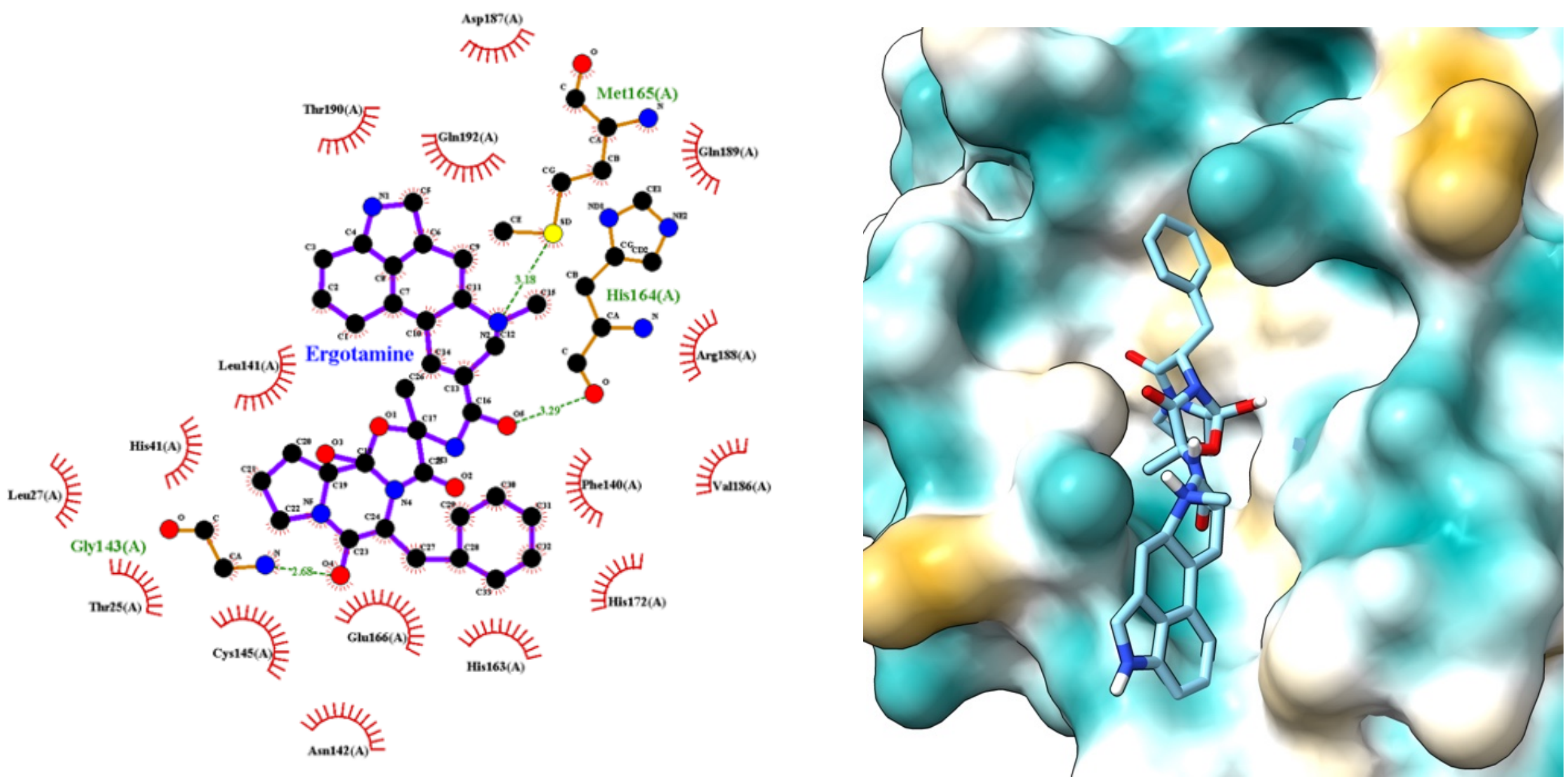

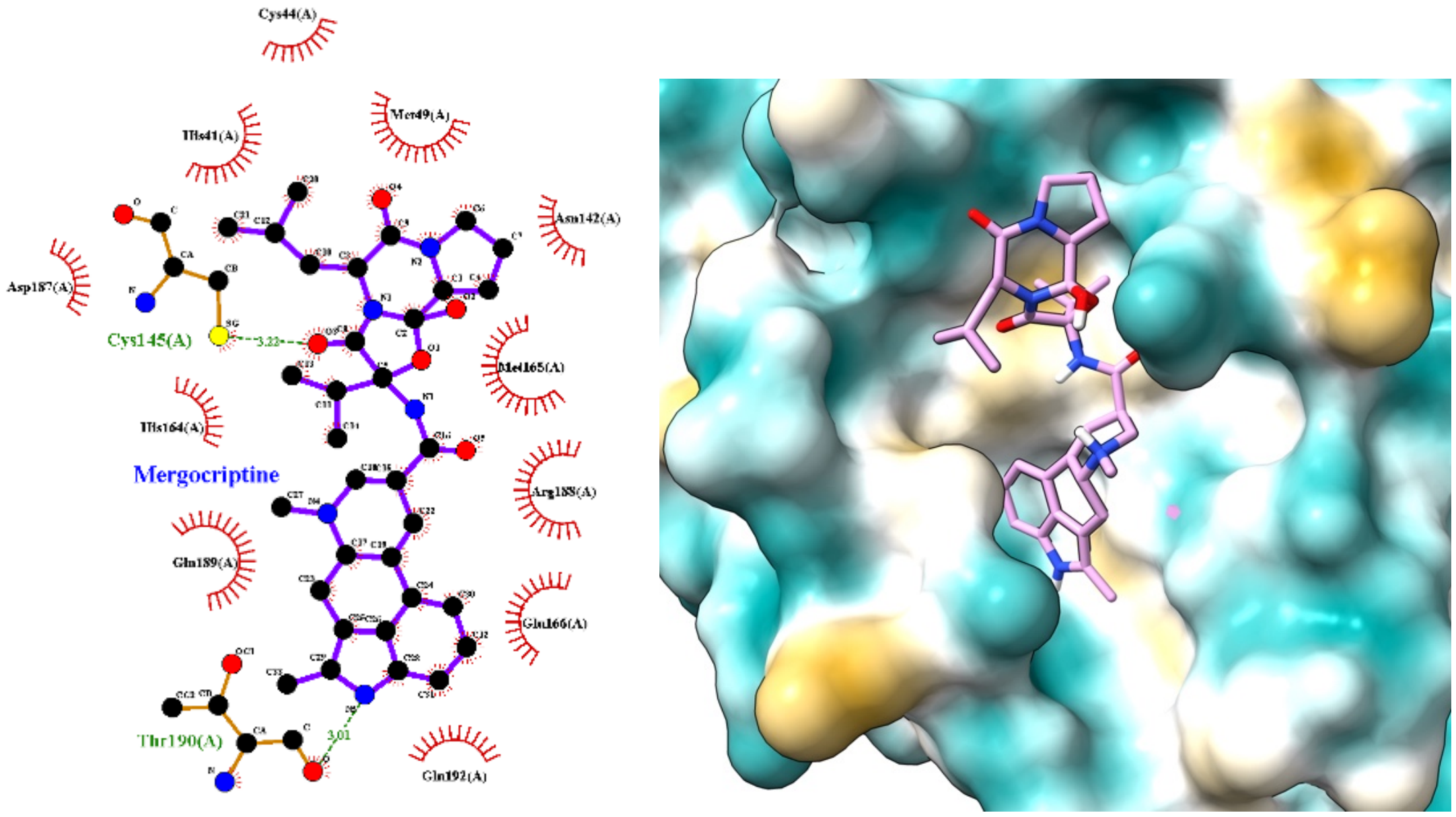

**Figure 3**. LigPlot (left) and hydrophobic protein surface representation (right) of the main interactions between M$^{pro}$ and ergotamine (top) and mergocriptine (bottom).

*Montelukast*

Montelukast is a cysteinyl leukotriene receptor antagonist used treat asthma and allergic rhinitis. It reduces pulmonary responses to antigen, tissue eosinophilia and IL-5 expression in inflammatory cells and decreases elevated levels of IL-1β and IL8 in viral upper respiratory tract infections.[35] Several computational have suggested putative binding to the terminal site of M$^{pro}$. Montelukast has been shown to inhibit at least one other protease, eosinophil protease.[36]

Mansoor and colleagues proposed that it could bind to M$^{pro}$ on the basis of a simple molecular docking study.[37] Wu et al also reported putative binding of montelukast to M$^{pro}$ in a computational study using the same Internal Coordinate Mechanics modelling methods.[38] No accurate binding affinities were reported in either study. Figure 4 shows a LigPlot representation of the interactions of key functional groups in montelukast with protease active site residues. These are also listed in Supplementary Table 2 for reference.

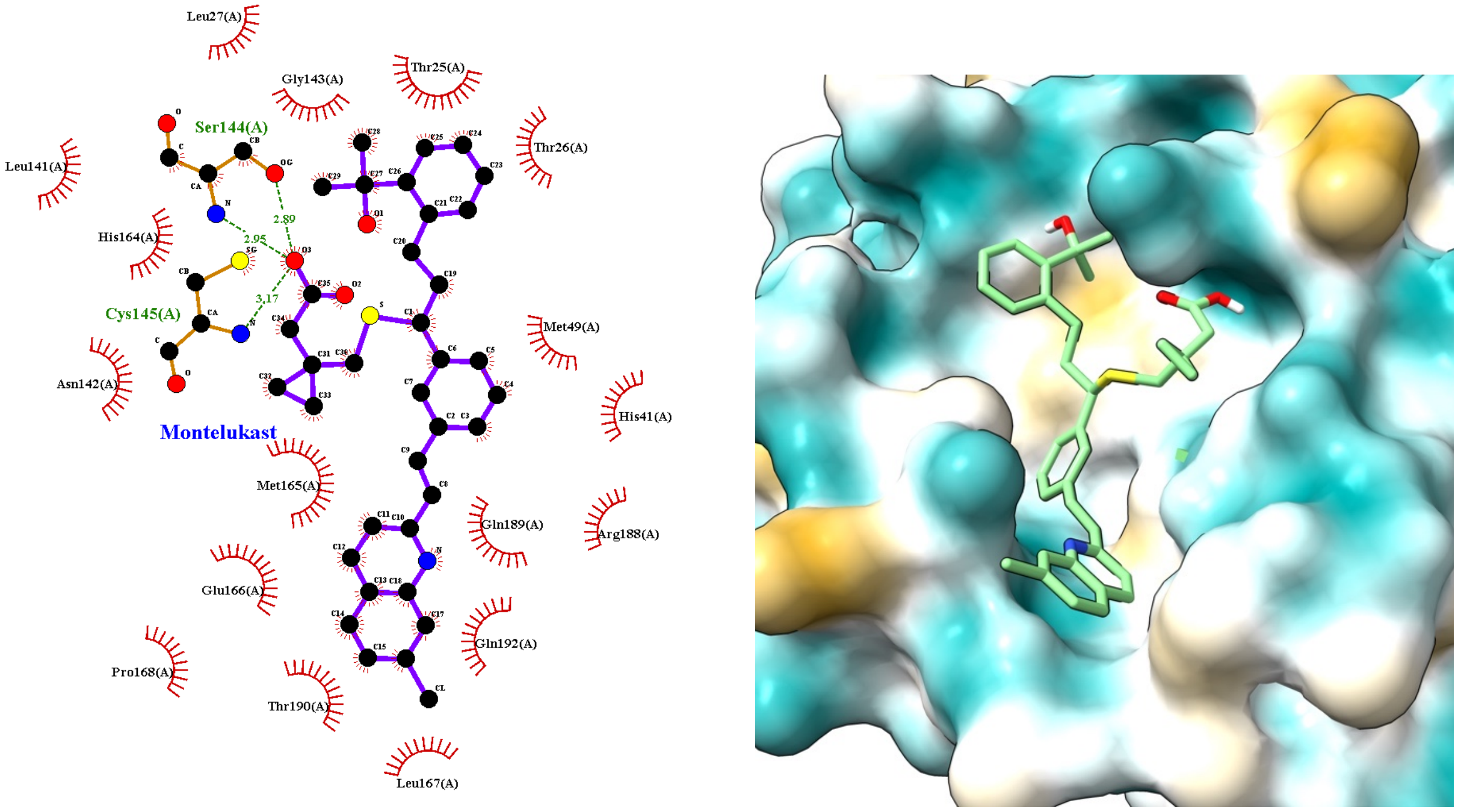

**Figure 4**. LigPlot (left) and hydrophobic protein surface representation (right) of the main interactions between $M^{pro}$ and montelukast.

*Bemcentinib*

Bemcentinib selectively inhibits AXL kinase activity, which blocks viral entry and enhances the antiviral type I interferon response. It's in vitro activity against SARS-Cov-2 has been assessed by several groups. In a Vero cell assay, Liu et al reported 10-40% protection at 50µM.[30] However, in an alternative assay using human Huh7.5 cells,[39] Bemcentinib exhibited an $IC_{50}$ of 100nM and $CC_{50}$ of 4.7µM. They also developed an assay in Vero cells and reported the $IC_{50}$ was 470nM and $CC_{50}$ was 1.6µM, considerably higher activity than that reported by Liu et al. As a result it is an investigational treatment for COVID-19 (www.clinicaltrialsregister.eu). Figure 5 shows a LigPlot representation of the interactions of key functional groups in bemcentinib with protease active site residues. These are also listed in Supplementary Table 2 for reference.

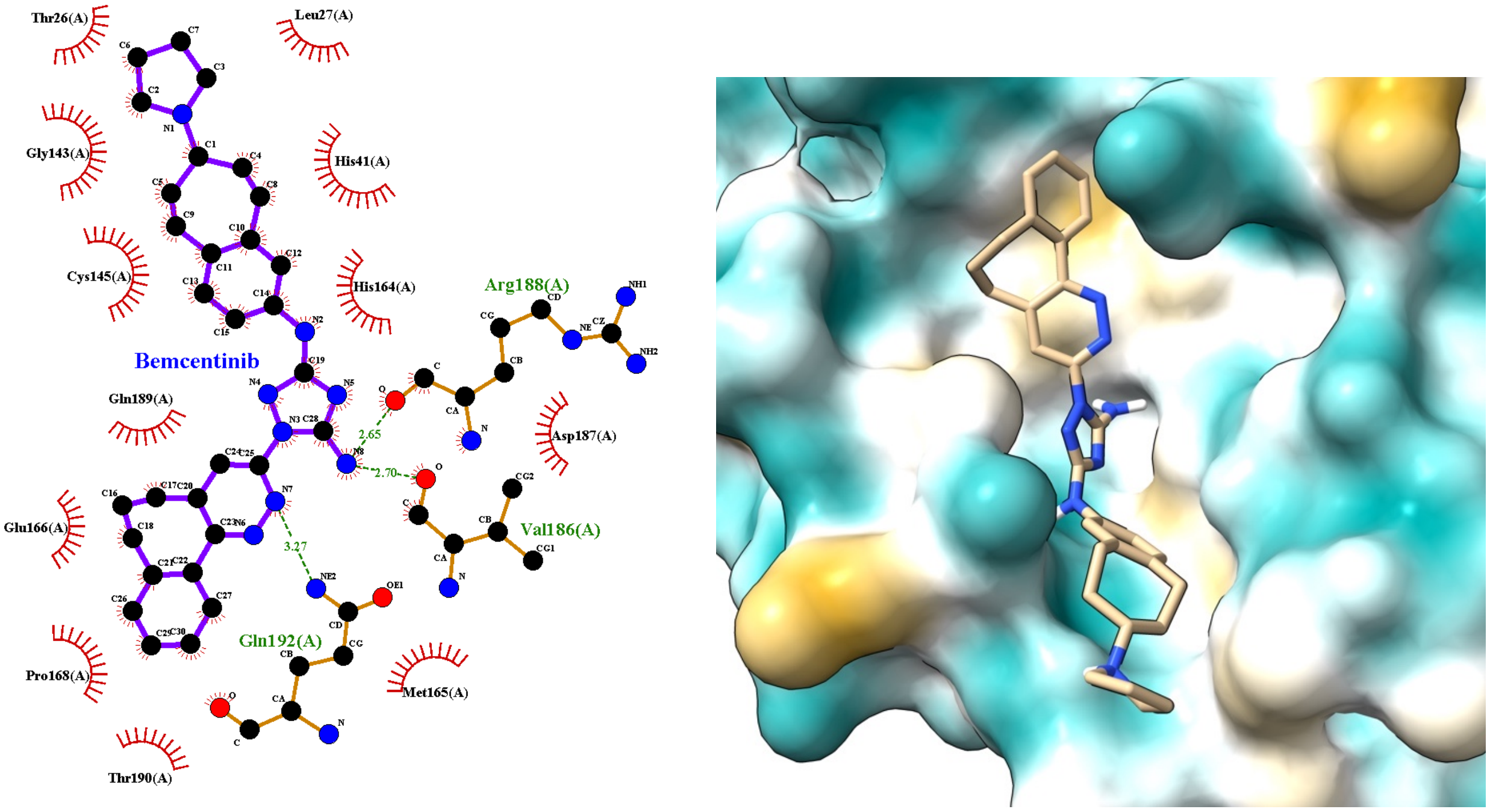

**Figure 5**. LigPlot (left) and hydrophobic protein surface representation (right) of the main interactions between M$^{pro}$ and bemcentinib.

*PC786*

PC786 targets the respiratory syncytial virus (RSV) L protein and is designed to be a topical inhalation treatment. There is very little published work on the SAR-Cov-2 efficacy or predicting binding affinity to M$^{pro}$. Panda and coworkers  reported a binding energy $\Delta G_{bind}$ of PC786 of −179.79, tighter binding than calculated for lopinavir (−131.49 kJ/mol), using a combined docking and MD approach.[40] Like our study, they employed Autodock Vina to dock a molecular library into the active site of M$^{pro}$, followed by MD simulation using GROMACS.  Figure 6 shows a LigPlot representation of the interactions of key functional groups in PC786 with protease active site residues. These are also listed in Supplementary Table 2 for reference.

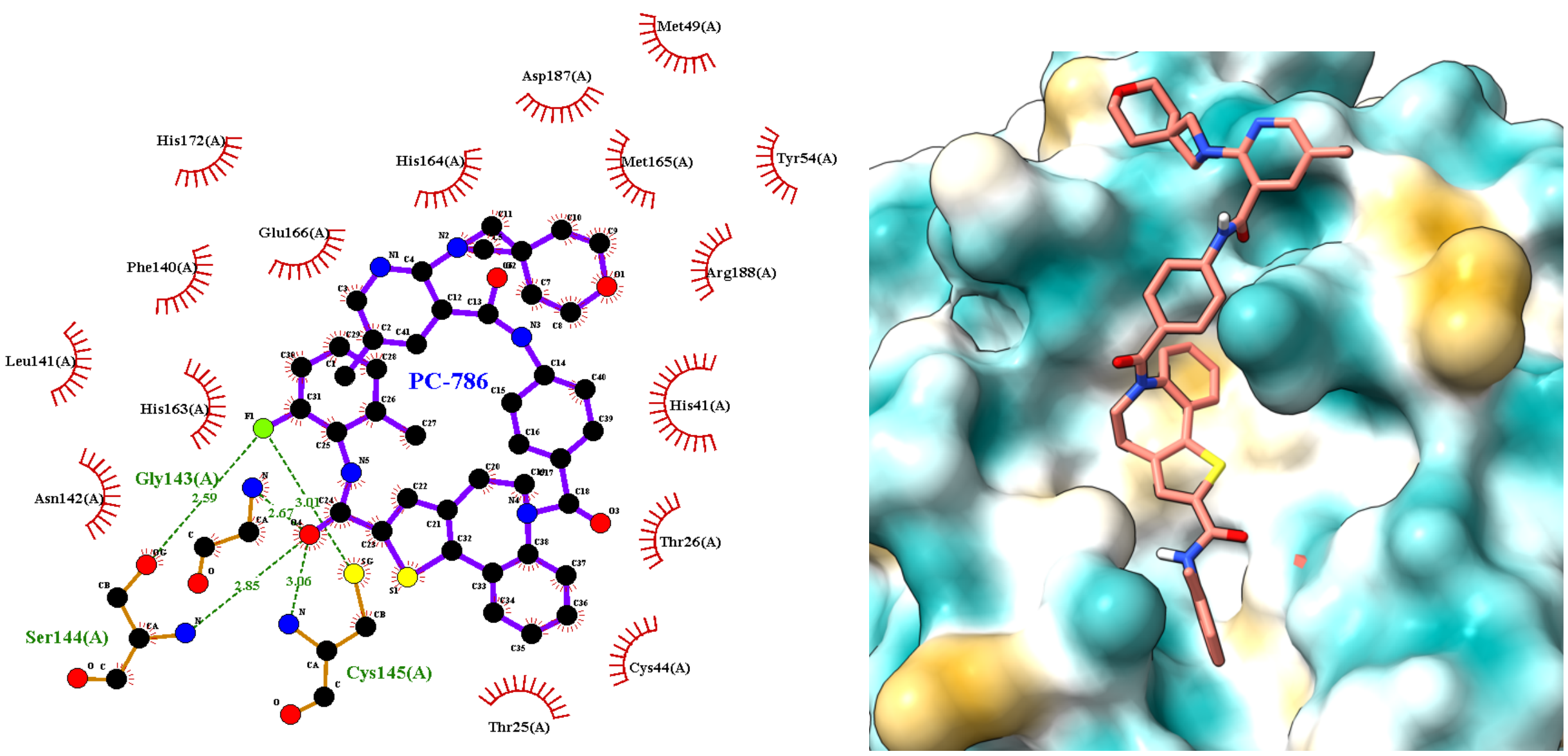

**Figure 6**. LigPlot (left) and hydrophobic protein surface representation (right) of the main interactions between $M^{pro}$ and PC786.

*Other novel putative $M^{pro}$ inhibitors from the short list of 84 drugs*

The predicted binding energies of the 84 drugs in the short list are summarized in Supplementary Table 1. We have also reviewed the literature for other in silico studies that have also identified some of these hit compounds as potential $M^{pro}$ inhibitors and have listed experimental in vitro and in vivo results and clinical trials in progress for drugs on the list. Two thirds of the drugs on the list have been reported to be potential inhibitors of SARS-Cov-2 target proteins, largely $M^{pro}$ but also RdRp, spike, helicase, human ACE2, 2'-O-methyltransferase nsp16/nsp10 complex, nsp1, $PL^{pro}$, nsp3, and nsp12. Satisfyingly, those with the best predicted binding affinity from our study have also been of greatest interest clinically, with more in vitro assay results and clinical trials for drugs with the highest binding affinities. This suggests that our screening and MD simulation methods are sufficiently robust and accurate to identify drugs for repurposing against SARS-Cov-2 and, more broadly, other coronaviruses. The 33% of drugs in the hit list that have no reported studies are therefore also of interest as novel drugs for COVID-19. We discuss some of the more interesting and novel hit compounds with higher binding affinities.

*Eltrombopag*

Eltrombopag is a TPO agonist that acts at the transmembrane domain of its cognate receptor C-Mpl via a histidine residue that occurs only in humans and apes. It scored highly in the docking studies, suggesting it could inhibit the 3CL protease and exhibit antiviral activity. Several other in silico screening studies also identified eltrombopag as a potential SARS-Cov-2 drug. Feng et al.'s studies suggested that eltrombopag bound not only to 3CL active site but also to the viral S-protein and to human ACE2.[41] This potential synergistic polypharmacy could be particularly beneficial.

Very little has been reported on the direct antiviral activity of eltrombopag. Recently, Vogel et al. reported direct inhibition of cytomegalovirus (CMV) by therapeutic doses of eltrombopag used to treat thrombocytopenia. [42] They showed that eltrombopag inhibits the late stages of the HCMV replication cycle and reduces virus titres by $1.8 \times 10^4$-fold at 10μM and by 15-fold at 500 nM. They suggested the mode of action was iron chelation and showed that eltrombopag was synergistic with ganciclovir in preventing viral replication.

Eltrombopag has also been proposed as a potential drug against SARS-CoV-2 spike protein on the basis of predicted strong binding to a pocket in the fusion cores of S2 domain.[41] Eltrombopag was also identified as a high binding affinity to human angiotensin converting enzyme 2 (ACE2), the primary binding site for the spike protein. Their virtual screen also used Autodock Vina, but no subsequent MD simulation was used for the top hit compounds from the screen. SPR was used to assess the binding of the drug to $M^{pro}$. Figure 7 shows a LigPlot representation of the interactions of key functional groups in eltrombopag with protease active site residues. These are also listed in Supplementary Table 2 for reference.

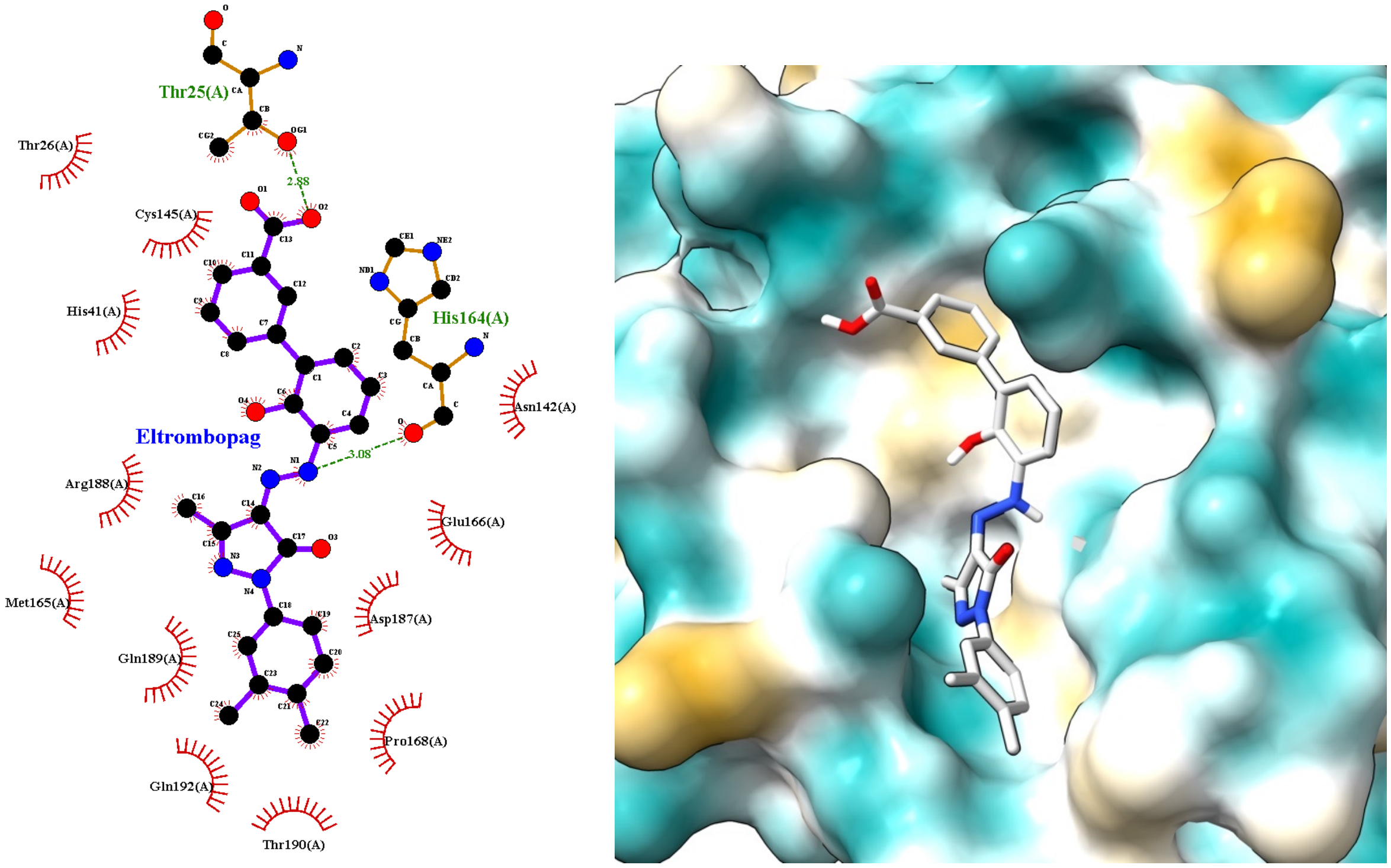

**Figure 7.** LigPlot (left) and hydrophobic M$^{pro}$ protein surface representation (right) of the main interactions between M$^{pro}$ and eltrombopag and M$^{pro}$.

Eltrombopag is of particular interest as a M$^{pro}$ inhibitor lead because it is novel and is also a member of a large class of small molecular TPO receptor agonists that may also exhibit activity against the viral protease, and potentially the spike protein and human ACE2.[43]. However, given the clotting disorders that SAR-Cov-2 generates, the TPOR agonist activities would need to be minimized to prevent platelet enhancement, while retaining or enhancing the antiviral activities.

Apart from the drugs discussed above, several other drugs in the list in Supplementary Table 1 are of interest. There are several other ergot derivatives with good predicted binding affinities to M$^{pro}$. Metergotamine and dihydroergocristine were predicted to have $\Delta G_{bind}$ of –29 and –24 kcal/mol respectively. Other drugs with binding energies stronger than –25 kcal/mol include galicaftor (clinical trial for cystic fibrosis), rolitetracycline (broad spectrum antibiotic), disogluside (natural product from *Dioscorea nipponica Makino* that reduces liver chronic inflammation and fibrosis), zafirlukast (leukotriene receptor antagonist for asthma), diosmin (a natural flavone for treating venous disease), AZD-5991 (clinical trial for relapsed or refractory haematologic malignancies), and ruzasvir (clinical trials for treatment of hepatitis C).

Li et al. has reported predicted $M^{pro}$ binding for galicaftor.[44] The protease binding of rolitetracycline has been reported by Durdagi et al. and Gul et al.[45, 46] Zhu and corkers measured the SARs-Cov-2 and $M^{pro}$ inhibition of zafirlukast.[47] The $IC_{50}$ for $M^{pro}$ was 24 µM and the $EC_{50}$ for the virus >20µM. The potential of the natural product diosmin as an antiviral agent targeting Mpro has also been reported in several recent computational studies.[48-51] Chakraborti et al. reported the potential of ruzasvir as a drug against SARS-Cov-2, although no data were provided.[52] These drugs and natural products merit assessment in SAR-Cov-2 assays and $M^{pro}$ inhibition experiments.

## Conclusions

Our virtual screening approach that applies Autodock Vina and MD simulation in tandem to calculate binding energies for repurposed drugs has identified 84 promising compounds for treating SARS-Cov2 infections. The screening was applied against the viral main protease $M^{pro}$ (3CLpro). The top hits from out study consisted of a mixture of antiviral agents , natural products and drugs that were developed for other applications and that have other models of action. The prognostic value of our computational approach has been demonstrated by the fact that it identified a diverse range of drugs that have been reported in other computational studies or that exhibit useful SARS-Cov-2 antiviral effects in vitro. The antiviral drugs simeprevir, sofosbuvir, lopinavir, ritonavir and remdesivir exhibit strong antiviral properties and several in in clinical trial or use against SARS-Cov-2. These drugs have been identified as binding to $M^{pro}$ also by numerous virtual screening studies and by in vitro assays. The more interesting and least studied lead drugs amongst our candidate list, bemcentinib, PC786, montelukast, ergotamine and mergocriptine, were predicted to have binding affinities equal to or greater than the antiviral drugs, and have also been shown to have *in vitro* antiviral activity against SARS-CoV-2. A few computational studies mostly using less rigorous methods than those we employed here, have also suggested that these drugs may bind to $M^{pro}$.

This high validation success rate strongly suggests that this type of virtual screening approach is capable of identifying compounds with potentially useful activity against SARS-CoV-2 and, by analogy, other coronaviruses. In particular, the 28 drugs for which no SARS-CoV-2 activity has been yet reported may be of particular interest for *in vitro* screening. The results of the current drug repurposing study provides information that could be useful to identify additional candidate drugs for testing for use in the current pandemic, as well as a rational computational paradigm for identifying therapeutic agents for future viral pandemics.

## Materials and Methods

*Protein structure preparation and grid preparation*

The crystal structure of the COVID-19 $M^{pro}$ (Figure 8) was downloaded from the RCSB PDB (http://www.rcsb.org; refcode 6Y2F).[18]

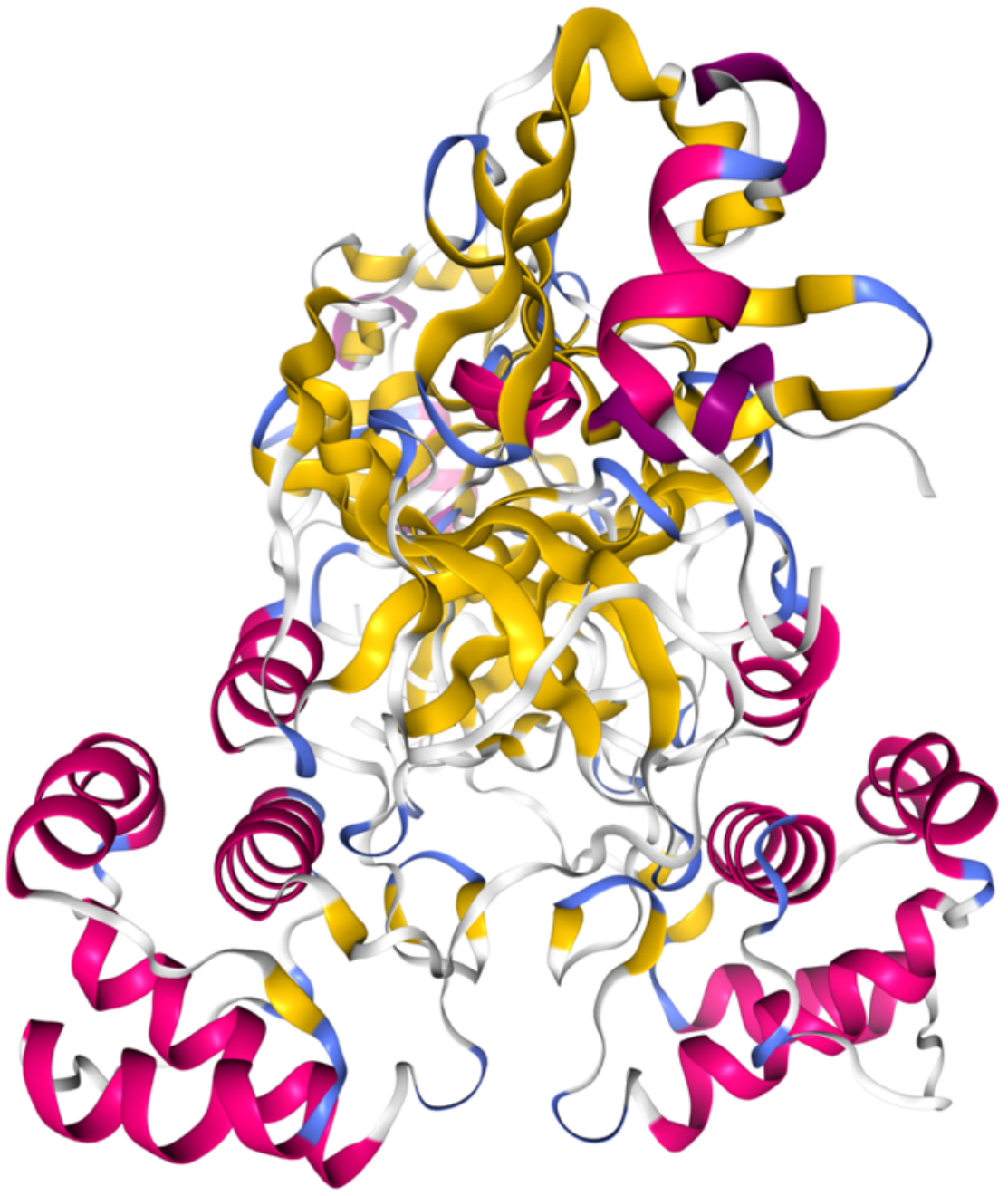

**Figure 8**. 3D structure of SARS-Cov-2 $M^{pro}$ (PDB refcode 6Y2F)

Protein preparation and removal of non-essential and non-bridging water molecules for docking studies were performed using the UCSF Chimera package (https://www.cgl.ucsf.edu/chimera/). [53] AutoDock Tools (ADT) software was used to prepare the required files for Autodock Vina by assigning hydrogen polarities, calculating Gasteiger charges to protein structures and converting protein structures from the .pdb file format to .pdbqt format.[54]. The surface area of the 3CLPro binding pocket is 335 Å$^2$, volume – 364.101 Å$^3$.[55]

*Screening databases*

Drugs database were downloaded from the DrugBank database (Wishart et al., 2018) and CHEMBL database (FDA approved) (Gaulton et al., 2017). A total of 8773 and 13,308 drugs were retrieved from Drugbank and CHEMBL database, respectively. The drugs were downloaded in sdf format and converted to .pdbqt format using Raccoon (Forli et al., 2016).

*Docking Methodology*

Small molecule ligand structures were docked against protein structure using the AutoDock Vina (version 1.1.3) package.[54] AutoDock Vina employs gradient-based conformational search approach and an energy-based empirical scoring function. AutoDock Vina is also flexible, easily scripted, extensively validated in many published studies with a variety of proteins and ligands and takes advantage of large multi-CPU or -GPU machines to run many calculations in parallel. The code has also been employed very successfully to dock millions of small molecule drug candidates into a series of protein targets to discover new potent drug leads. The package includes useful scripts for generating modified .pdb files required for grid calculations and for setting up the grid calculations around each protein automatically. The software requires the removal of hydrogens, addition of polar hydrogens, setting of the correct atom types, and calculation of atom charges compatible with the AutoGrid code. The algorithm generates a grid around each protein and calculates the interaction energy of a probe noble gas atom at each grid position outside and within internal cavities of the protein. The grid resolution was set to 1 Å, the maximum number of binding modes to output was fixed at 10, and the exhaustiveness level (controlling the number of independent runs performed) was set at 8. The docking employed a genetic algorithm to optimize the binding conformations of the ligands during docking to the protease site. Drugs were docked individually to the active site of $M^{pro}$ (3CLPro, refcode 6Y2F) with the grid coordinates (grid centre) and grid boxes of appropriate sizes generated by the bash script vina_screen.sh (Supplementary Information). The top scored compounds were identified with a python script script1.py (Supplementary Information) and subjected to molecular dynamic simulation. The docked structures were analysed using UCSF Chimera [53] and LigPlot+ software[56] to illustrate hydrogen-bond and hydrophobic interactions. A total of fifty top compounds selected from each of the Drugbank and CHEMBL compounds. Sixteen compounds were common to both database top hits. Molecular dynamics studies were conducted on the unique set of eighty-four compounds from both sets.

*Molecular Dynamics Simulation*

The top screened compound complexes with protease were minimized with CHARMm force field. The topology files of the ligands were prepared from Swissparam (http://www.swissparam.ch/) [57] and minimized in Gromacs2020 (http://www.gromacs.org/).[58]. Docked complexes of ligands and COVID-19 $M^{pro}$ protein were used as starting geometries for MD simulations. Simulations were carried out using the GPU accelerated version of the program with the CHARMm force field I periodic boundary conditions in ORACLE server. Docked complexes were immersed in a truncated octahedron box of TIP3P water molecules. The solvated box was further neutralized with Na+ or Cl− counter ions using the tleap program. Particle Mesh Ewald (PME) was employed to calculate the long-range electrostatic interactions. The cut-off distance for the long-range van der Waals (VDW) energy term was 12.0 Å. The whole system was minimized without any restraint. The above steps applied 2500 cycles of steepest descent minimization followed by 5000 cycles of conjugate gradient minimization. After system optimization, the MD simulations was initiated by gradually heating each system in the NVT ensemble from 0 to 300 K for 50 ps using a Langevin thermostat with a coupling coefficient of 1.0/ps and with a force constant of 2.0 kcal/mol·Å2 on the complex. Finally, a production run of 20 ns of MD simulation was performed under a constant temperature of 300 K in the NPT ensemble with periodic boundary conditions for each system. During the MD procedure, the SHAKE algorithm was applied for the constraint of all covalent bonds involving hydrogen atoms. The time step was set to 2 fs. The structural stability of the complex was monitored by the RMSD and RMSF values of the backbone atoms of the entire protein. Calculations were also performed for up to 100 ns on few compounds to ensure that 20ns is sufficiently long for convergence. Duplicate production runs starting with different random seeds were also run to allow estimates of binding energy uncertainties to be determined.

The binding free energies of the protein-protein complexes were evaluated in two ways. The traditional method is to calculate the energies of solvated SARS-Cov-2 protease and small molecule ligands and that of the bound complex and derive the binding energy by subtraction.

$$\Delta G \text{ (binding, aq)} = \Delta G \text{ (complex, aq)} - (\Delta G \text{ (protein, aq)} + \Delta G \text{ (ligand, aq)} \quad (1)$$

We also calculated binding energies using the molecular mechanics Poisson Boltzmann surface area (MM/PBSA) tool in GROMACS that is derived from the nonbonded interaction energies of the complex. The method is also widely used method for binding free energy calculations.

MMPBSA calculations were conducted by GMXPBSA 2.1[59] a suite based on Bash/Perl scripts for streamlining MM/PBSA calculations on structural ensembles derived from GROMACS trajectories and to automatically calculate binding free energies for protein–protein or ligand–protein. GMXPBSA 2.1 calculates diverse MM/PBSA energy contributions from molecular mechanics (MM) and electrostatic contribution to solvation (PB) and non-polar contribution to solvation (SA). This tool combines the capability of MD simulations (GROMACS) and the Poisson–Boltzmann equation (APBS) for calculating solvation energy (Baker et., 2001). The g_mmpbsa tool in GROMACS was used after molecular dynamics simulations, the output files obtained were used to post-process binding free energies by the single-trajectory MMPBSA method. In the current study we considered 100 frames at equal distance from 20ns trajectory files.

Specifically, for a non-covalent binding interaction in the aqueous phase the binding free energy, $\Delta G$ (bind,aq), is: –

$$\Delta G\ (bind,aqu) = \Delta G\ (bind,vac) + \Delta G\ (bind,solv) \quad (2)$$

where ΔG (bind,vac) is the binding free energy in vacuum, and ΔG(bind,solv) is the solvation free energy change upon binding: –

$$\Delta G\ (bind,solv) = \Delta G\ (R{:}L,\ solv) - \Delta G\ (R,solv) - \Delta G\ (L,solv) \quad (3)$$

where ΔG (R:L,solv), ΔG (R,solv) and ΔG (L,solv) are solvation free energies of complex, receptor and ligand, respectively.

While this manuscript was in preparation, Guterres and Im showed how substantial improvement in protein-ligand docking results could be achieved using high-throughput MD simulations.[60] As with our study, they also employed AutoDock Vina for docking, followed by MD simulation using CHARMM. The MD parameters they advocated were very similar to those used in our study. Proteins were solvated in a box of TIP3P water molecules extending 10 Å beyond the proteins and the particle-mesh Ewald method was used for electrostatic interactions. Nonbonded interactions over 10 and 12 Å were truncated. Their systems were minimized for 5000 steps using the steepest descent method followed by 1 ns equilibration with an NVT setting. For each protein-ligand complex, they ran 3 × 100 ns production runs from the same initial structure using different initial velocity random seeds and an integration step size of 2 fs. Over 56 protein targets (of 7 different protein classes) and 560 ligands this show 22% improvement in the area under receiver operating characteristics curve, from an initial value of 0.68 using AutoDock Vina alone to a final value of 0.83 when the Vina results were refined by MD.

**Acknowledgements**

We would like to thank Harinda Rajapaksha for assistance to optimise GROMACS for this project. We would also like to thank Oracle for providing their Cloud computing resources for the modelling studies described herein. In particular, we wish to thank Peter Winn, Dennis Ward, and Alison Derbenwick Miller from Oracle in facilitating these studies. The opinions expressed herein are solely those of the individual authors and should not be inferred to reflect the views of their affiliated institutions, funding bodies or Oracle corporation.

**Author contributions**

Petrovsky - conceived project, analysed data, contributed to manuscript; Piplani and Kumar Singh - performed the computations, analysed data, contributed to the manuscript; Winkler - analysed data and contributed to manuscript

## References

1. Zhang, J.; Zeng, H.; Gu, J.; Li, H.; Zheng, L.; Zou, Q., Progress and Prospects on Vaccine Development against SARS-CoV-2. *Vaccines* **2020,** *8* (2), 153.

2. Whitworth, J., COVID-19: a fast-evolving pandemic. *Trans. R. Soc. Trop. Med. Hyg.* **2020,** *114* (4), 241-248.

3. Thanh Le, T.; Andreadakis, Z.; Kumar, A.; Gomez Roman, R.; Tollefsen, S.; Saville, M.; Mayhew, S., The COVID-19 vaccine development landscape. *Nat. Rev. Drug Discov.* **2020,** *19*, 305-306.

4. Sohrabi, C.; Alsafi, Z.; O'Neill, N.; Khan, M.; Kerwan, A.; Al-Jabir, A.; Iosifidis, C.; Agha, R., World Health Organization declares global emergency: A review of the 2019 novel coronavirus (COVID-19). *Int. J. Surg.* **2020,** *76*, 71-76.

5. Schlagenhauf, P.; Grobusch, M. P.; Maier, J. D.; Gautret, P., Repurposing antimalarials and other drugs for COVID-19. *Travel Med. Infect. Dis.* **2020**, 101658.

6. Sanders, J. M.; Monogue, M. L.; Jodlowski, T. Z.; Cutrell, J. B., Pharmacologic Treatments for Coronavirus Disease 2019 (COVID-19): A Review. *J. Am. Med. Assoc.* **2020,** *323*, 1824-1836.

7. Rosales-Mendoza, S.; Marquez-Escobar, V. A.; Gonzalez-Ortega, O.; Nieto-Gomez, R.; Arevalo-Villalobos, J. I., What Does Plant-Based Vaccine Technology Offer to the Fight against COVID-19? *Vaccines* **2020,** *8* (2).

8. Rosa, S. G. V.; Santos, W. C., Clinical trials on drug repositioning for COVID-19 treatment. *Rev. Panam. Salud. Publica* **2020,** *44*, e40.

9. Olsen, M.; Cook, S. E.; Huang, V.; Pedersen, N.; Murphy, B. G., Perspectives: Potential Therapeutic Options for SARS-CoV-2 Patients Based on Feline Infectious Peritonitis Strategies: Central Nervous System Invasion and Drug Coverage. *Int. J. Antimicrob. Agents* **2020**, 105964.

10. Mendes, A., Research towards treating COVID-19. *Br. J. Commun. Nurs.* **2020,** *25* (4), 204-205.

11. Lu, S., Timely development of vaccines against SARS-CoV-2. *Emerg. Microbes Infect.* **2020,** *9* (1), 542-544.

12. Jiang, S., Don't rush to deploy COVID-19 vaccines and drugs without sufficient safety guarantees. *Nature* **2020,** *579* (7799), 321.

13. Ciotti, M.; Angeletti, S.; Minieri, M.; Giovannetti, M.; Benvenuto, D.; Pascarella, S.; Sagnelli, C.; Bianchi, M.; Bernardini, S.; Ciccozzi, M., COVID-19 Outbreak: An Overview. *Chemother.* **2020**, 1-9.

14. Berkley, S., COVID-19 needs a big science approach. *Science* **2020,** *367*, 1407.

15. Zhang, J.-J.; Shen, X.; Yan, Y.-M.; Yan, W.; Cheng, Y.-X. Discovery of anti-SARS-CoV-2 agents from commercially available flavor via docking screening *OSF Preprints* [Online], 2020. osf.io/vjch2.

16. Zhang, T.; He, Y.; Xu, W.; Ma, A.; Yang, Y.; Xu, K. F., Clinical trials for the treatment of Coronavirus disease 2019 (COVID-19): A rapid response to urgent need. *Sci. China Life Sci.* **2020,** *63* (5), 774-776.

17. Yavuz, S.; Unal, S., Antiviral Treatment of Covid-19. *Turk. J. Med. Sci.* **2020,** *50* ((SI-1)), 611-619.

18. Zhang, L.; Lin, D.; Sun, X.; Curth, U.; Drosten, C.; Sauerhering, L.; Becker, S.; Rox, K.; Hilgenfeld, R., Crystal structure of SARS-CoV-2 main protease provides a basis for design of improved α-ketoamide inhibitors. *Science* **2020,** *368*, 409-12.

19. Dai, W.; Zhang, B.; Jiang, X. M.; Su, H.; Li, J.; Zhao, Y.; Xie, X.; Jin, Z.; Peng, J.; Liu, F.; Li, C.; Li, Y.; Bai, F.; Wang, H.; Cheng, X.; Cen, X.; Hu, S.; Yang, X.; Wang, J.; Liu, X.; Xiao, G.; Jiang, H.; Rao, Z.; Zhang, L. K.; Xu, Y.; Yang, H.; Liu, H., Structure-based design of antiviral drug candidates targeting the SARS-CoV-2 main protease. *Science* **2020,** *368* (6497), 1331-1335.

20. Saul, S.; Einav, S., Old Drugs for a New Virus: Repurposed Approaches for Combating COVID-19. *ACS Infect. Dis.* **2020**.

21. Rathnayake, A. D.; Zheng, J.; Kim, Y.; Perera, K. D.; Mackin, S.; Meyerholz, D. K.; Kashipathy, M. M.; Battaile, K. P.; Lovell, S.; Perlman, S.; Groutas, W. C.; Chang, K., 3C-like protease inhibitors block coronavirus replication in vitro and improve survival in MERS-CoV–infected mice. *Sci. Transl. Med.* **2020,** *12* (557), eabc5332.

22. Mengist, H. M.; Fan, X.; Jin, T., Designing of improved drugs for COVID-19: Crystal structure of SARS-CoV-2 main protease M(pro). *Signal Transduct. Target Ther.* **2020,** *5* (1), 67.

23. Abhithaj, J.; Dileep, F.; Sharanya, C. S.; Arun, K. G.; Sadasivan, C.; E., J. V. Repurposing Simeprevir, Calpain Inhibitor IV and a Cathepsin F Inhibitor Against SARS-CoV-2: A Study Using in Silico Pharmacophore Modeling and Docking Methods *ChemRxiv* [Online], 2020. chemrxiv.12317213.

24. Lo, H. S.; Hui, K. P.; Lai, H.-M.; Khan, K. S.; Kaur, S.; Li, Z.; Chan, A. K.; Cheung, H. H.-Y.; Ng, K. C.; Ho, J. C. W. Simeprevir suppresses SARS-CoV-2 replication and synergizes with remdesivir *bioRxiv* [Online], 2020. 2020.05.26.116020.

25. Bolcato, G.; Bissaro, M.; Pavan, M.; Sturlese, M.; Moro, S., Targeting the Coronavirus SARS-CoV-2: computational insights into the mechanism of action of the protease inhibitors Lopinavir, Ritonavir, and Nelfinavir. *Sci. Rep.* **2020,** *under consideration*.

26. Costanzo, M.; De Giglio, M. A. R.; Roviello, G. N., SARS-CoV-2: recent reports on antiviral therapies based on lopinavir/ritonavir, darunavir/umifenovir, hydroxychloroquine, remdesivir, favipiravir and other drugs for the treatment of the new coronavirus. *Curr. Med. Chem.* **2020,** *27*, 4536-4541.

27. Muralidharan, N.; Sakthivel, R.; Velmurugan, D.; Gromiha, M. M., Computational studies of drug repurposing and synergism of lopinavir, oseltamivir and ritonavir binding with SARS-CoV-2 Protease against COVID-19. *J. Biomol. Struct. Dynam.* **2020**, 1-6.

28. Hendaus, M. A., Remdesivir in the treatment of Coronavirus Disease 2019 (COVID-19): A simplified summary. *J. Biomol. Struct. Dynam.* **2020**, Accepted.

29. Al-Khafaji, K.; Al-DuhaidahawiL, D.; Taskin Tok, T., Using integrated computational approaches to identify safe and rapid treatment for SARS-CoV-2. *J. Biomol. Struct. Dynam.* **2020**, Accepted.

30. Liu, S.; Lien, C. Z.; Selvaraj, P.; Wang, T. T. Evaluation of 19 antiviral drugs against SARS-CoV-2 Infection *bioRxiv* [Online], 2020. 2020.04.29.067983.

31. Ma, C.; Sacco, M. D.; Hurst, B.; Townsend, J. A.; Hu, Y.; Szeto, T.; Zhang, X.; Tarbet, B.; Marty, M. T.; Chen, Y.; Wang, J., Boceprevir, GC-376, and calpain inhibitors II, XII inhibit SARS-CoV-2 viral replication by targeting the viral main protease. *Cell Res.* **2020,** *30*, 678–692.

32. Gurung, A. B.; Ali, M. A.; Lee, J.; Farah, M. A.; Al-Anazi, K. M., In silico screening of FDA approved drugs reveals ergotamine and dihydroergotamine as potential coronavirus main protease enzyme inhibitors. *Saudi J. Biol. Sci.* **2020**, Accepted.

33. Mevada, V.; Dudhagara, P.; Gandhi, H.; Vaghamshi, N.; Beladiya, U.; Patel, R. Drug repurposing of approved drugs Elbasvir, Ledipasvir, Paritaprevir, Velpatasvir, Antrafenine and Ergotamine for combating COVID19 *ChemRxiv* [Online], 2020. chemrxiv.12115251.v2.

34. Qiao, Z.; Zhang, H.; Ji, H. F.; Chen, Q., Computational View toward the Inhibition of SARS-CoV-2 Spike Glycoprotein and the 3CL Protease. *Computat.* **2020,** *8* (2), 53.

35. Almerie, M. Q.; Kerrigan, D. D., The association between obesity and poor outcome after COVID-19 indicates a potential therapeutic role for montelukast. *Med. Hypoth.* **2020,** *143*, 109883.

36. Langlois, A.; Ferland, C.; Tremblay, G. M.; Laviolette, M., Montelukast regulates eosinophil protease activity through a leukotriene-independent mechanism. *J. Allergy Clin. Immunol.* **2006,** *118* (1), 113-119.

37. Mansoor, S.; Saadat, S.; Amin, A.; Ali, I.; Ghaffar, M. T.; Amin, U.; Mukhtar, M. A Case for Montelukast in COVID-19: The use of Computational Docking to estimate the effects of Montelukast on potential viral main protease catalytic site *Res. Square* [Online], 2020. rs.3.rs-27079/v1.

38. Wu, C.; Liu, Y.; Yang, Y.; Zhang, P.; Zhong, W.; Wang, Y.; Wang, Q.; Xu, Y.; Li, M.; Li, X., Analysis of therapeutic targets for SARS-CoV-2 and discovery of potential drugs by computational methods. *Acta Pharmaceut. Sin. B* **2020,** *10* (5), 766-788.

39. Dittmar, M.; Lee, J. S.; Whig, K.; Segrist, E.; Li, M.; Jurado, K.; Samby, K.; Ramage, H.; Schultz, D.; Cherry, S. Drug repurposing screens reveal FDA approved drugs active against SARS-Cov-2 *bioRxiv* [Online], 2020. 2020.06.19.161042.

40. Panda, P. K.; Arul, M. N.; Patel, P.; Verma, S. K.; Luo, W.; Rubahn, H.-G.; Mishra, Y. K.; Suar, M.; Ahuja, R., Structure-based drug designing and immunoinformatics approach for SARS-CoV-2. *Science Adv.* **2020,** *6*, eabb8097.

41. Feng, S.; Luan, X.; Wang, Y.; Wang, H.; Zhang, Z.; Wang, Y.; Tian, Z.; Liu, M.; Xiao, Y.; Zhao, Y.; Zhou, R.; Zhang, S., Eltrombopag is a potential target for drug intervention in SARS-CoV-2 spike protein. *Infect. Genet. Evol.* **2020,** *85*, 104419.

42. Vogel, J. U.; Schmidt, S.; Schmidt, D.; Rothweiler, F.; Koch, B.; Baer, P.; Rabenau, H.; Michel, D.; Stamminger, T.; Michaelis, M.; Cinatl, J., The Thrombopoietin Receptor Agonist Eltrombopag Inhibits Human Cytomegalovirus Replication Via Iron Chelation. *Cells* **2019,** *9* (1), 31.

43. Tarasova, A.; Winkler, D. A., Modelling atypical small-molecule mimics of an important stem cell cytokine, thrombopoietin. *ChemMedChem* **2009,** *4* (12), 2002-11.

44. Li, Y.; Zhang, Y.; Han, Y.; Zhang, T.; Du, R. Prioritization of Potential Drugs Targeting the SARS-CoV-2 Main Protease *ChemRxiv* [Online], 2020. chemrxiv.12629858.v1.

45. DURDAĞI, S., Virtual drug repurposing study against SARS-CoV-2 TMPRSS2 target. *Turk. J. Biol.* **2020**, 44.

46. Gul, S.; Ozcan, O.; Okyar, A.; Barıs, I.; Kavakli, I. H., In Silico Identification of Widely Used and Well Tolerated Drugs That May Inhibit SARSCov-2 3C-like Protease and Viral RNA-Dependent RNA Polymerase Activities, and May Have Potential to Be Directly Used in Clinical Trials. *J. Biomol. Struct. Dynam.* **2020**.

47. Zhu, W.; Xu, M.; Chen, C. Z.; Guo, H.; Shen, M.; Hu, X.; Shinn, P.; Klumpp-Thomas, C.; Michael, S. G.; Zheng, W. Identification of SARS-CoV-2 3CL Protease Inhibitors by a Quantitative High-throughput Screening *bioRxiv* [Online], 2020. 2020.07.17.207019.

48. Arun, K.; Sharanya, C.; Abhithaj, J.; Francis, D.; Sadasivan, C., Drug repurposing against SARS-CoV-2 using E-pharmacophore based virtual screening, molecular docking and molecular dynamics with main protease as the target. *J. Biomol. Struct. Dynam.* **2020**, 1-12.

49. Ngo, S. T.; Pham, N. Q. A.; Le, L.; Pham, D.-H.; Vu, V., Computational Determination of Potential Inhibitors of SARS-CoV-2 Main Protease. *J. Chem. Inf. Mod.* **2020**, Accepted.

50. Peterson, L., In Silico Molecular Dynamics Docking of Drugs to the Inhibitory Active Site of SARS-CoV-2 Protease and Their Predicted Toxicology and ADME. *SSRN 3580951* **2020**.

51. Peterson, L. COVID-19 and Flavonoids: In Silico Molecular Dynamics Docking to the Active Catalytic Site of SARS-CoV and SARS-CoV-2 Main Protease *SSRN* [Online], **2020**. 3599426.

52. Chakraborti, S.; Bheemireddy, S.; Srinivasan, N. Repurposing drugs against main protease of SARS-CoV-2: mechanism based insights supported by available laboratory and clinical data 2020. chemrxiv.12057846.v2.

53. Pettersen, E. F.; Goddard, T. D.; Huang, C. C.; Couch, G. S.; Greenblatt, D. M.; Meng, E. C.; Ferrin, T. E., UCSF Chimera--a visualization system for exploratory research and analysis. *J. Comp. Chem.* **2004,** *25* (13), 1605-12.

54. Forli, S.; Huey, R.; Pique, M. E.; Sanner, M. F.; Goodsell, D. S.; Olson, A. J., Computational protein-ligand docking and virtual drug screening with the AutoDock suite. *Nat Protoc* **2016,** *11* (5), 905-19.

55. Tian, W.; Chen, C.; Lei, X.; Zhao, J.; Liang, J., CASTp 3.0: computed atlas of surface topography of proteins. *Nucleic Acids Res.* **2018,** *46* (W1), W363-W367.

56. Laskowski, R. A.; Swindells, M. B., LigPlot+: multiple ligand-protein interaction diagrams for drug discovery. *J. Chem. Inf. Model.* **2011,** *51* (10), 2778-86.

57. Zoete, V.; Cuendet, M. A.; Grosdidier, A.; Michielin, O., SwissParam: a fast force field generation tool for small organic molecules. *J. Comp. Chem.* **2011,** *32* (11), 2359-68.

58. Abraham, M. J.; Murtola, T.; Schulz, R.; Páll, S.; Smith, J. C.; Hessa, B.; Lindahlad, E., GROMACS; High performance molecular simulations through multi-level parallelism from laptops to supercomputers. *SoftwareX* **2015,** *1-2*, 19-25.

59. Paissoni, C.; Spiliotopoulos, D.; Musco, G.; Spitaleri, A., GMXPBSA 2.1: A GROMACS tool to perform MM/PBSA and computational alanine scanning. *Comp. Phys. Comm.* **2015,** *186*, 105–107.

60. Guterres, H.; Im, W., Improving Protein-Ligand Docking Results with High-Throughput Molecular Dynamics Simulations. *J. Chem. Inf .Model.* **2020,** *60* (4), 2189-2198.

**Computational screening of repurposed drugs and natural products against SARS-Cov-2 main protease ($M^{pro}$ ) as potential COVID-19 therapies**

Sakshi Pilani[1-2], Puneet Singh[2], Nikolai Petrovsky[1-2], David A. Winkler[3-6]

[1] College of Medicine and Public Health, Flinders University, Bedford Park 5046, Australia

[2] Vaxine Pty Ltd, 11 Walkley Avenue, Warradale 5046, Australia

[3] La Trobe University, Kingsbury Drive, Bundoora 3042, Australia

[4] Monash Institute of Pharmaceutical Sciences, Monash University, Parkville 3052, Australia

[5] School of Pharmacy, University of Nottingham, Nottingham NG7 2RD. UK

[6] CSIRO Data61, Pullenvale 4069, Australia

## Supplementary information

**Table S1**. Binding energies and published SARS-Cov-2 data for 84 top ranked small molecule ligands

| | Name | ChEMBL (C)) or Drugbank (D) ID | $\Delta G_{MMPBSA}$ kcal/mol | SARS-Cov-2 data |
|---|---|---|---|---|
| 1 | Bemcentinib | C 3809489 | -33.9 | Phase 2 clinical trial,[1] $ED_{50}$ 0.1 (Huh7.5), 0.47 (Vero), 2.1 (Calu3) µM,[2] predicted 2'-O-methyltransferase nsp16/nsp10 complex binding[3] |
| 2 | PC786 | C 4291143 | -33.0 | Predicted spike glycoprotein,[4] $M^{pro}$, and ACE2 binding[5] |
| 3 | Montelukast | C 787 | -32.6 | Significant reduction in SARS-CoV-2 infection in elderly asthmatic patients treated with MK.[6] Several predicted $M^{pro}$ binding studies e.g.[6, 7] |
| 4 | Ergotamine | C 442 | -31.6 | Several predicted $M^{pro}$ binding studies e.g.[8-10] |
| 5 | Simeprevir | D06290 | -31.5 | In vitro $EC_{50}$ 4.08 µM, and many predicted $M^{pro}$ binding studies e.g.[11], many predicted Mpro binding studies e.g.[12-14], predicted RdRp binding[15, 16] |
| 6 | Sofosbuvir | D08934 | -31.0 | In vitro $EC_{50}$ values of 6.2 and 9.5 µM.[17] Predicted RdRp[18, 19] binding |
| 7 | Lopinavir | D01601 | -30.7 | In vitro $EC_{50}$ 5.73 µM,[20] Multiple single agent and combination human trials e.g.[21, 22]. In vitro $EC_{50}$ 26.63 µM.[23] Predicted $M^{pro}$ binding[24-26] |
| 8 | Ritonavir | D00503 | -30.2 | In vitro $EC_{50}$ 8.63 µM,[20] Multiple single agent and combination human trials e.g.[21, 27] Predicted $M^{pro}$ [26] helicase and RdRp binding[28] |
| 9 | Mergocriptine | C2105887 | -30.1 | Predicted 2′-O-ribose methyltransferase[29] binding |
| 10 | Remdesivir | D14761 | -29.9 | Multiple human trials e.g.[30, 31], in vitro $EC_{50}$ 23.15 µM,[23] predicted $M^{pro}$ [32] and RdRp[33] binding |
| 11 | Metergotamine | C2106428 | -29.1 | Predicted 2′-O-ribose methyltransferase[29] binding |

| | Name | ChEMBL (C)) or Drugbank (D) ID | $\Delta G_{MMPBSA}$ kcal/mol | SARS-Cov-2 data |
|---|---|---|---|---|
| 12 | Galicaftor | D14894 | -28.4 | Predicted $M^{pro}$ binding[34] |
| 13 | Eltrombopag | C461101 | -28.2 | Predicted spike[35] and RdRp[8] binding |
| 14 | Saquinavir | C114 | -27.9 | In vitro $EC_{50}$ 8.83 μM,[20] predicted ACE2, $M^{pro}$,[10] and RdRp[36] binding[37] |
| 15 | Rolitetracycline | C 1237046 | -27.6 | Predicted $M^{pro}$ [38] and spike[39] binding |
| 16 | Disogluside | C395414 | -27.3 | Predicted $M^{pro}$ [29] binding |
| 17 | Zafirlukast | D00549 | -26.7 | Predicted spike,[10] $M^{pro}$,[40] and 2′-O-ribose methyltransferase[41] binding |
| 18 | Diosmin | D08995 | -25.8 | Predicted $M^{pro}$ [42, 43] binding |
| 19 | AZD-5991 | D14792 | -25.2 | … |
| 20 | Ruzasvir | C 3833385 | -25.1 | Predicted $M^{pro}$ [44] and RdRp[16] binding |
| 21 | Rebastinib | C1738757 | -24.3 | Predicted 2'-O-ribose methyltransferase nsp16 binding[29] |
| 22 | RSV-604 | D15197 | -24.3 | … |
| 23 | Eravacycline | D12329 | -24.2 | Predicted $M^{pro}$ [45, 46] binding |
| 24 | Lifitegrast | C2048028 | -24.1 | Predicted nsp16/nsp10 complex[3] and RdRp[16] binding |
| 25 | 10-Deoxymethynolide | D07703 | -24.1 | … |
| 26 | Ledipasvir | D09027 | -23.8 | Predicted $M^{pro}$ binding[47] |
| 27 | Deldeprevir | C3040582 | -23.8 | Predicted $M^{pro}$ binding[13] |
| 28 | Rifamycin | D11753 | -23.8 | … |
| 29 | Ethoxazorutoside | C 2106047 | -23.8 | … |
| 30 | Dihydroergocristine | C601773 | -23.8 | Predicted $M^{pro}$ [47] binding |
| 31 | Gedatolisib | C592445 | -23.8 | … |
| 32 | Lorecivivint | D14883 | -23.8 | Predicted $M^{pro}$ [42] and spike[5] binding |
| 33 | MK-6325 | C4297304 | -23.6 | … |
| 34 | Laniquidar | C539378 | -23.1 | Predicted RdRp[48] and spike[49] binding |
| 35 | Tirabrutinib | C 4071161 | -22.9 | $EC_{50}$ >10μM in SARS-CoV-2-Nluc neutralization assay[50] |
| 36 | 3-(2-aminoquinazolin-6-yl)-4-methyl-N-[3- | D06925 | -22.9 | … |

| | Name | ChEMBL (C)) or Drugbank (D) ID | ΔG$_{MMPBSA}$ kcal/mol | SARS-Cov-2 data |
|---|---|---|---|---|
| | (trifluoromethyl)phenyl]benzamide | | | |
| 37 | Ensartinib | D14860 | -22.8 | … |
| 38 | Anacetrapib | C1800807 | -22.4 | … |
| 39 | Pazinaclone | C2107504 | -22.3 | Predicted M$^{pro}$ binding[51] |
| 40 | BMS-986142 | D15291 | -22.2 | … |
| 41 | Phthalocyanine | D12983 | -22.2 | Predicted nsp1,[52] M$^{pro}$,[34] spike,[49] nsp1,[52] and 2′-O-methyltransferase[3] binding |
| 42 | Umbralisib | C3948730 | -21.9 | … |
| 43 | DNK333 | C105060 | -21.9 | … |
| 44 | Midostaurin | C608533 | -21.7 | Predicted M$^{pro}$ binding.[40] |
| 45 | Umifenovir | D13609 | -21.6 | Multiple clinical trials only show higher –ve rate of PCR on day 14 in adult COVID-19 patients.[53] Shorten the viral shedding interval.[54] Predicted M$^{pro}$ binding.[55] |
| 46 | Lumacaftor | D09280 | -21.5 | Predicted M$^{pro}$ [47, 48, 56] and RdRp[57] binding |
| 47 | TU-100 | D12467 | -21.3 | … |
| 48 | Triamcinolone furetonide | C2105791 | -21.2 | … |
| 49 | Zoliflodacin | C3544978 | -21.2 | Predicted M$^{pro}$ [58] and PL$^{pro}$[48] binding |
| 50 | KPT-9274 | C4297467 | -20.9 | … |
| 51 | Atazanavir | D01072 | -20.8 | Inhibits SARS-CoV-2 replication, predicted M$^{pro}$ [59] and helicase[60] binding |
| 52 | Mitratapide | C2104975 | -20.8 | Predicted 2'-O-ribose methyl-transferase nsp16 binding[29] |
| 53 | Tarloxotinib | D14944 | -20.5 | … |
| 54 | Spergualin | C1765508 | -20.5 | … |
| 55 | Moxidectin | D11431 | -20.5 | … |
| 56 | PRI-724 | D15034 | -20.0 | … |
| 57 | 2-[3-(methyl[1-(2-naphthoyl)piperidin-4-yl] amino}carbonyl)-2-naphthyl]-1-(1-naphthyl)-2-oxoethyl phosphonic | D04016 | -19.8 | … |

| | Name | ChEMBL (C)) or Drugbank (D) ID | $\Delta G_{MMPBSA}$ kcal/mol | SARS-Cov-2 data |
|---|---|---|---|---|
| | acid | | | |
| 58 | ASP-4058 | D11819 | -19.8 | … |
| 59 | Beclabuvir | DB12225 | -19.7 | Predicted RdRp[15] and $M^{pro}$ [61] binding |
| 60 | Ubrogepant | C2364638 | -19.5 | Predicted to disrupt spike-ACE2 interaction[62] |
| 61 | Dihydroergotamine | D00320 | -19.5 | Predicted $M^{pro}$ [8, 9] and 2′-O-ribose methyltransferase[41] binding |
| 62 | Lifirafenib | C 4209157 | -19.3 | Predicted spike[48] and nsp10– nsp16 complex[3] binding |
| 63 | Golvatinib | D11977 | -18.8 | Predicted $M^{pro}$ [48] and RdRp[36] binding |
| 64 | Tirilazad | D13050 | -18.6 | Predicted $M^{pro}$ [25, 61] and nsp1[52] binding |
| 65 | 4-[(10s,14s,18s)-18-(2-amino-2-oxoethyl)-14-(1-naphthylmethyl)-8,17,20-trioxo-7,16,19-triaza spiro[5.14]icos-11-en-10-yl]benzylphosphonic acid | D03276 | -18.6 | … |
| 66 | Etamocycline | C3989417 | -16.1 | |
| 67 | Quarfloxin | C3989407 | -15.9 | Predicted spike,[49] $PL^{pro}$[48] and $M^{pro}$ [37] |
| 68 | 2'-(4-dimethylamino-phenyl)-5-(4-methyl-1-piperazinyl)-2,5'-bi-benzimidazole | D04011 | -15.7 | … |
| 69 | Dihydrostreptomycin | C 1950576 | -15.6 | Predicted nsp3 and nsp10– nsp16 complex binding[63] |
| 70 | Rimegepant | C2178422 | -15.6 | … |
| 71 | Bezitramide | C2104149 | -15.4 | … |
| 72 | Flutroline | C57241 | -15.3 | Predicted 2'-O-ribose methyltransferase nsp16 binding[29] |
| 73 | Carfilzomib | D08889 | -15.2 | Predicted $M^{pro}$ binding[45] |
| 74 | IPI-549 | C3984425 | -15.0 | … |
| 75 | Milademetan | C4292264 | -14.6 | Predicted $M^{pro}$ binding[45] |
| 76 | Nemiralisib | C2216859 | -14.4 | Predicted $M^{pro}$ binding[64] |
| 77 | Amrubicin | C1186894 | -14.3 | Predicted $M^{pro}$ [65] binding |
| 78 | Genz-10850 | D04289 | -13.1 | Predicted nsp12 binding[66] |

| | Name | ChEMBL (C)) or Drugbank (D) ID | $\Delta G_{MMPBSA}$ kcal/mol | SARS-Cov-2 data |
|---|---|---|---|---|
| 79 | Penimepicycline | C 3833378 | -12.9 | Predicted to $M^{pro}$ and spike[38] binding |
| 80 | Tipifarnib | C289228 | -12.6 | … |
| 81 | MK3207 | C1910936 | -12.1 | Predicted $M^{pro}$,[67] PLpro,[68] 2'-O-ribose methyl-transferase nsp16[29] binding |
| 82 | Naldemedine | C2105791 | -12.1 | Predicted $M^{pro}$ [40] and spike RBD [69] binding |
| 83 | Tariquidar | D06240 | -12.0 | … |
| 84 | Netupitant | C206253 | -11.9 | Predicted RdRp binding [14] and $M^{pro}$ [40] |

**Table S2**. Binding interactions with $M^{pro}$ binding site for top 10 ranked drugs.

| ID | Drug | Interacting Residues | H-Bond |
|---|---|---|---|
| CHEMBL7835 | Bemcitinib | THR26, LEU27, HIS41, ASN142, GLY143, SER144, CYS145, HIS164, MET165, GLU166, LEU167, PRO168, VAL186, ASP187, ARG188, GLN189, THR190, ALA191, GLN192 | VAL186 (O-N8) 2.70<br>ARG188(O-N8) 2.65<br>GLN192 (NE2 -N7) 3.27 |
| CHEMBL442 | Ergotamine | THR25, LEU27, HIS41, PHE140, LEU141, ASN142, GLY143, SER144, CYS145, HIS163, HIS164, MET165, GLU166, HIS172, VAL186, ASP187, ARG188, GLN189, THR190, GLN192 | GLY143(N-O4) 2.68<br>HIS164 (O-O5) 3.29<br>MET165(SD-C12) 3.18 |
| DB01601 | Lopinavir | THR26, HIS41, MET49, PHE140, LEU141, ASN142, CYS145, HIS163, HIS164, MET165, GLU166, HIS172, VAL186, ASP187, ARG188, GLN189, THR190, GLN192 | ASN142 (OD1-O2) 2.59 |
| CHEMBL4958 | Mergocriptine | HIS41, CYS44, MET49, ASN142, GLY143, SER144, CYS145, HIS163, HIS164, MET165, GLU166, LEU167, PRO168, VAL186, ASP187, ARG188, GLN189, THR190, GLN192 | CYS145(SG-O3) 3.22<br>THR190 (O-N5) 3.01 |
| CHEMBL4291143 | PC-786 | THR25, THR26, HIS41, CYS44, MET49, TYR54, PHE140, LEU141, ASN142, GLY143, SER144, CYS145, HIS163, HIS164, MET165, GLU166, HIS172, ASP187, ARG188, GLN189 | GLY143(N-O4) 2.67<br>SER144(OG-F1) 2.59<br>SER144(N-O4) 2.85<br>CYS145 (SG-F1) 3.01<br>CYS145 (N-)4) 3.06 |
| DB14761 | Remdesivir | HIS41, MET49, PHE140, LEU141, ASN142, GLY143, SER144, CYS145, HIS163, HIS164, MET165, GLU166, HIS172, VAL186, ASP187, ARG188, THR190, GLN192 | PHE140(O-N5) 2.98<br>SER144(OG-N6) 3.14<br>HIS163(NE2-N6) 3.01<br>HIS164(O-O4) 2.67 |

| ID | Drug | Interacting Residues | H-Bond |
|---|---|---|---|
| DB00503 | Ritonavir | THR25, THR26, LEU27, HIS41, CYS44, THR45, SER46, MET47, PHE140, LEU141, ASN142, GLY143, SER144, CYS145, HIS163, HIS164, MET165, GLU166, LEU167, PRO168, HIS172, VAL186, ARG188, GLN189, THR190, GLN192 | CYS145(SG-O3) 3.03<br>HIS164(O-O3) 2.88 |
| DB06290 | Simeprevir | HIS41, CYS44, MET49, TYR54, PHE140, LEU141, ASN142, GLY143, SER144, CYS145, HIS163, HIS164, MET165, GLU166, LEU167, PRO168, THR169, GLY170, HIS172, VAL186, ARG188, GLN189, THR190, GLN192 | HIS163(NE2-O4) 3.09<br>HIS164(O-N3) 3.18<br>CYS145(O-HO) 3.34 |
| DB08934 | Sofosbuvir | HIS41, MET49, TYR54, PHE140, LEU141, ASN142, GLY143, SER144, CYS145, HIS163, HIS164, MET165, GLU166, LEU167, PRO168, HIS172, VAL186, ARG188, GLN189, THR190, GLN192, ALA193 | SER144(OG-O9) 3.09<br>GLU166(N-O6) 3.28 |
| CHEMBL4499 | Montelukast | THR25, THR26, LEU27, HIS41, MET49, TYR54, PHE140, LEU141, ASN142, GLY143, SER144, CYS145, HIS163, HIS164, MET165, GLU166, LEU167, PRO168, ASP187, ARG188, GLN189, THR190, GLN192 | SER144 (N-O3) 2.95<br>SER144 (OG-O3) 2.89<br>CYS145 (N-O3) 3.17 |

**Scripts**:

**1)Conf.txt**

```
receptor = 6Y2F.pdbqt

center_x=  9.245

center_y=  -0.788

center_z = 18.371

size_x = 50

size_y = 50

size_z = 50

num_modes = 10

exhaustiveness = 50
```

**2)vina_screen.sh**

```
#! /bin/bash

for f in CHEMBL*.pdbqt; do

   b=`basename $f .pdbqt`

   echo Processing ligand $b

mkdir -p $b

   vina --config conf.txt --cpu 50 --ligand $f --out $[b]/out.pdbqt --log $[b]/log.txt

done
```

**3)Script1.py**

```
#! /usr/bin/env python

import sys
```

```
import glob
def doit(n):
file_names = glob.glob('*/*.pdbqt')
    everything = []
    failures = []
    print 'Found', len(file_names), 'pdbqt files'
    for file_name in file_names:
        file = open(file_name)
        lines = file.readlines()
file.close()
        try:
            line = lines[1]
            result = float(line.split(':')[1].split()[0])
everything.append([result, file_name])
        except:
failures.append(file_name)
everything.sort(lambda x,y: cmp(x[0], y[0]))
    part = everything[:n]
    for p in part:
        print p[1],
    print
    if len(failures) > 0:
        print 'WARNING:', len(failures), 'pdbqt files could not be processed'
if __name__ == '__main__':
doit(int(sys.argv[1]))
```

## References

1. Wilkinson, T.; Dixon, R.; Page, C.; Carroll, M.; Griffiths, G.; Ho, L. P.; De Soyza, A.; Felton, T.; Lewis, K. E.; Phekoo, K.; Chalmers, J. D.; Gordon, A.; McGarvey, L.; Doherty, J.; Read, R. C.; Shankar-Hari, M.; Martinez-Alier, N.; O'Kelly, M.; Duncan, G.; Walles, R.; Sykes, J.; Summers, C.; Singh, D.; Collaborators, A., ACCORD: A Multicentre, Seamless, Phase 2 Adaptive Randomisation Platform Study to Assess the Efficacy and Safety of Multiple Candidate Agents for the Treatment of COVID-19 in Hospitalised Patients: A structured summary of a study protocol for a randomised controlled trial. *Trials* **2020,** *21* (1), 691.

2. Dittmar, M.; Lee, J. S.; Whig, K.; Segrist, E.; Li, M.; Jurado, K.; Samby, K.; Ramage, H.; Schultz, D.; Cherry, S. Drug repurposing screens reveal FDA approved drugs active against SARS-Cov-2 *bioRxiv* [Online], 2020. 2020.06.19.161042.

3. Encinar, J. A.; Menendez, J. A., Potential Drugs Targeting Early Innate Immune Evasion of SARS-Coronavirus 2 via 2'-O-Methylation of Viral RNA. *Viruses* **2020,** *12* (5).

4. Allam, A. E.; Assaf, H. K.; Hassan, H. A.; Shimizu, K.; Elshaier, Y. A. M. M., An in silico perception for newly isolated flavonoids from peach fruit as privileged avenue for a countermeasure outbreak of COVID-19. *RSC Adv.,* **2020,** *10*, 29983-29998.

5. Panda, P. K.; Arul, M. N.; Patel, P.; Verma, S. K.; Luo, W.; Rubahn, H.-G.; Mishra, Y. K.; Suar, M.; Ahuja, R., Structure-based drug designing and immunoinformatics approach for SARS-CoV-2. *Sci. Adv.* **2020,** *6*, eabb8097.

6. Bozek, A.; Winterstein, J., Montelukast's ability to fight COVID-19 infection. *J. Asthma* **2020**, 1-2.

7. Copertino, D. C.; Duarte, R. R. R.; Powell, T. R.; de Mulder Rougvie, M.; Nixon, D. F., Montelukast drug activity and potential against severe acute respiratory syndrome coronavirus 2 (SARS-CoV-2). *J. Med. Virol.* **2020**.

8. Gul, S.; Ozcan, O.; Okyar, A.; Barıs, I.; Kavakli, I. H., In Silico Identification of Widely Used and Well Tolerated Drugs That May Inhibit SARSCov-2 3C-like Protease and Viral RNA-Dependent RNA Polymerase Activities, and May Have Potential to Be Directly Used in Clinical Trials. *J. Biomol. Struct. Dynam.* **2020**.

9. Gurung, A. B.; Ali, M. A.; Lee, J.; Farah, M. A.; Al-Anazi, K. M., In silico screening of FDA approved drugs reveals ergotamine and dihydroergotamine as potential coronavirus main protease enzyme inhibitors. *Saudi J. Biol. Sci.* **2020**, Accepted.

10. Qiao, Z.; Zhang, H.; Ji, H. F.; Chen, Q., Computational View toward the Inhibition of SARS-CoV-2 Spike Glycoprotein and the 3CL Protease. *Computat.* **2020,** *8* (2), 53.

11. Lo, H. S.; Hui, K. P.; Lai, H.-M.; Khan, K. S.; Kaur, S.; Li, Z.; Chan, A. K.; Cheung, H. H.-Y.; Ng, K. C.; Ho, J. C. W. Simeprevir suppresses SARS-CoV-2 replication and synergizes with remdesivir *bioRxiv* [Online], 2020. 2020.05.26.116020.

12. Alamri, M. A.; Tahir ul Qamar, M.; Mirza, M. U.; Bhadane, R.; Alqahtani, S. M.; Muneer, I.; Froeyen, M.; Salo-Ahen, O. M., Pharmacoinformatics and molecular dynamics simulation studies reveal potential covalent and FDA-approved inhibitors of SARS-CoV-2 main protease 3CLpro. *J. Biomol. Struct. Dynam.* **2020**, 1-13.

13. Hakmi, M.; Bouricha, E.; Kandoussi, I.; El Harti, J.; Ibrahimi, A., Repurposing of known anti-virals as potential inhibitors for SARS-CoV-2 main protease using molecular docking analysis. *Bioinformat.* **2020,** *16* (4), 301-305.

14. Hosseini, M.; Chen, W.; Wang, C. Computational Molecular Docking and Virtual Screening Revealed Promising SARS-CoV-2 Drugs. *ChemRxiv* [Online], 2020. chemrxiv.12237995.v1.

15. Beg, M. A.; Athar, F., Anti-HIV and Anti-HCV drugs are the putative inhibitors of RNA-dependent-RNA polymerase activity of NSP12 of the SARS CoV- 2 (COVID-19). *Pharm. Pharmacol. Int. J.* **2020,** *8* (3), 163–172.

16. Cozac, R.; Medzhidov, N.; Yuk, S. Predicting inhibitors for SARS-CoV-2 RNA-dependent RNA polymerase using machine learning and virtual screening *arXiv* [Online], 2020. arXiv:2006.06523.

17. Sacramento, C.; Fintelman-Rodrigues, N.; Temerozo, J. R.; da Silva Gomes Dias, S.; Ferreira, A. C.; Mattos, M.; Pão, C. R. R.; de Freitas, C. S.; Soares, V. C.; Bozza, F. A.; Bou-Habib, D. C.; Bozza, P. T.; T.M.L., S. The in vitro antiviral activity of the anti-hepatitis C virus (HCV) drugs daclatasvir and sofosbuvir against SARS-CoV-2 *bioRxiv* [Online], 2020. 2020.06.15.153411.

18. Jacome, R.; Campillo-Balderas, J. A.; Ponce de Leon, S.; Becerra, A.; Lazcano, A., Sofosbuvir as a potential alternative to treat the SARS-CoV-2 epidemic. *Sci. Rep.* **2020,** *10* (1), 9294.

19. Elfiky, A.; Ibrahim, N.; Elshemey, W., Drug repurposing against MERS CoV and SARS-COV-2 PLpro in silico. *Research Square* rs.3.rs-19600/v1 **2020**. https://doi.org/10.21203/rs.3.rs-19600/v1

20. Yamamoto, N.; Matsuyama, S.; Hoshino, T.; Yamamoto, N. Nelfinavir inhibits replication of severe acute respiratory syndrome coronavirus 2 in vitro *bioRxiv* [Online], 2020. 2020.04.06.026476.

21. Cao, B.; Wang, Y.; Wen, D.; Liu, W.; Wang, J.; Fan, G.; Ruan, L.; Song, B.; Cai, Y.; Wei, M.; Li, X.; Xia, J.; Chen, N.; Xiang, J.; Yu, T.; Bai, T.; Xie, X.; Zhang, L.; Li, C.; Yuan, Y.; Chen, H.; Li, H.; Huang, H.; Tu, S.; Gong, F.; Liu, Y.; Wei, Y.; Dong, C.; Zhou, F.; Gu, X.; Xu, J.; Liu, Z.; Zhang, Y.; Li, H.; Shang, L.; Wang, K.; Li, K.; Zhou, X.; Dong, X.; Qu, Z.; Lu, S.; Hu, X.; Ruan, S.; Luo, S.; Wu, J.; Peng, L.; Cheng, F.; Pan, L.; Zou, J.; Jia, C.; Wang, J.; Liu, X.; Wang, S.; Wu, X.; Ge, Q.; He, J.; Zhan, H.; Qiu, F.; Guo, L.; Huang, C.; Jaki, T.; Hayden, F. G.; Horby, P. W.; Zhang, D.; Wang, C., A Trial of Lopinavir-Ritonavir in Adults Hospitalized with Severe Covid-19. *N Engl J Med* **2020,** *382* (19), 1787-1799.

22. Costanzo, M.; De Giglio, M. A. R.; Roviello, G. N., SARS-CoV-2: recent reports on antiviral therapies based on lopinavir/ritonavir, darunavir/umifenovir, hydroxychloroquine, remdesivir, favipiravir and other drugs for the treatment of the new coronavirus. *Curr. Med. Chem.* **2020,** *27*, 4536-4541.

23. Choy, K. T.; Wong, A. Y.; Kaewpreedee, P.; Sia, S. F.; Chen, D.; Hui, K. P. Y.; Chu, D. K. W.; Chan, M. C. W.; Cheung, P. P.; Huang, X.; Peiris, M.; Yen, H. L., Remdesivir, lopinavir, emetine, and homoharringtonine inhibit SARS-CoV-2 replication in vitro. *Antiviral Res* **2020,** *178*, 104786.

24. Bolcato, G.; Bissaro, M.; Pavan, M.; Sturlese, M.; Moro, S., Targeting the Coronavirus SARS-CoV-2: computational insights into the mechanism of action of the protease inhibitors Lopinavir, Ritonavir, and Nelfinavir. *Scientif. Rep.* **2020,** *under consideration*.

25. Chtita, S.; Belhassan, A.; Aouidate, A.; Belaidi, S.; Bouachrine, M.; Lakhlifi, T., Discovery of Potent SARS-CoV-2 Inhibitors from Approved Antiviral Drugs via Docking Screening. *Comb Chem High Throughput Screen* **2020**.

26. Muralidharan, N.; Sakthivel, R.; Velmurugan, D.; Gromiha, M. M., Computational studies of drug repurposing and synergism of lopinavir, oseltamivir and ritonavir binding with SARS-CoV-2 Protease against COVID-19. *J. Biomol. Struct. Dynam.* **2020**, 1-6.

27. Verdugo-Paiva, F.; Izcovich, A.; Ragusa, M.; Rada, G., Lopinavir-ritonavir for COVID-19: A living systematic review. *Medwave* **2020,** *20* (6), e7967.

28. Beck, B. R.; Shin, B.; Choi, Y.; Park, S.; Kang, K., Predicting commercially available antiviral drugs that may act on the novel coronavirus (SARS-CoV-2) through a drug-target interaction deep learning model. *Comput. Struct. Biotechnol. J.* **2020,** *18*, 784-790.

29. Jiang, Y.; Liu, L.; Manning, M.; Bonahoom, M.; Lotvola, A.; Yang, Z.-Q. Repurposing Therapeutics to Identify Novel Inhibitors Targeting 2'-O-Ribose Methyltransferase Nsp16 of SARS-CoV-2 *ChemRxiv* [Online], 2020. chemrxiv.12252965.v1.

30. Olender, S. A.; Perez, K. K.; Go, A. S.; Balani, B.; Price-Haywood, E. G.; Shah, N. S.; Wang, S.; Walunas, T. L.; Swaminathan, S.; Slim, J.; Chin, B.; De Wit, S.; Ali, S. M.; Soriano Viladomiu, A.; Robinson, P.; Gottlieb, R. L.; Tsang, T. Y. O.; Lee, I. H.; Haubrich, R. H.; Chokkalingam, A. P.; Lin, L.; Zhong, L.; Bekele, B. N.; Mera-Giler, R.; Gallant, J.; Smith, L. E.; Osinusi, A. O.; Brainard, D. M.; Hu, H.; Phulpin, C.; Edgar, H.; Diaz-Cuervo, H.; Bernardino, J. I., Remdesivir for Severe COVID-19 versus a Cohort Receiving Standard of Care. *Clin Infect Dis* **2020**.

31. Wang, Y.; Zhang, D.; Du, G.; Du, R.; Zhao, J.; Jin, Y.; Fu, S.; Gao, L.; Cheng, Z.; Lu, Q.; Hu, Y.; Luo, G.; Wang, K.; Lu, Y.; Li, H.; Wang, S.; Ruan, S.; Yang, C.; Mei, C.; Wang, Y.; Ding, D.; Wu, F.; Tang, X.; Ye, X.; Ye, Y.; Liu, B.; Yang, J.; Yin, W.; Wang, A.; Fan, G.; Zhou, F.; Liu, Z.; Gu, X.; Xu, J.; Shang, L.; Zhang, Y.; Cao, L.; Guo, T.; Wan, Y.; Qin, H.; Jiang, Y.; Jaki, T.; Hayden, F. G.; Horby, P. W.; Cao, B.; Wang, C., Remdesivir in adults with severe COVID-19: a randomised, double-blind, placebo-controlled, multicentre trial. *Lancet* **2020,** *395* (10236), 1569-1578.

32. Rehman, M. T.; AlAjmi, M. F.; Hussain, A. Natural Compounds as Inhibitors of SARS-CoV-2 Main Protease (3CLpro): A Molecular Docking and Simulation Approach to Combat COVID-19. *ChemRxiv* [Online], 2020. chemrxiv.12362333.v2.

33. Elfiky, A. A., Ribavirin, Remdesivir, Sofosbuvir, Galidesivir, and Tenofovir against SARS-CoV-2 RNA dependent RNA polymerase (RdRp): A molecular docking study. *Life Sci* **2020,** *253*, 117592.

34. Li, Y.; Zhang, Y.; Han, Y.; Zhang, T.; Du, R. Prioritization of Potential Drugs Targeting the SARS-CoV-2 Main Protease *ChemRxiv* [Online], 2020. chemrxiv.12629858.v1.

35. Feng, S.; Luan, X.; Wang, Y.; Wang, H.; Zhang, Z.; Wang, Y.; Tian, Z.; Liu, M.; Xiao, Y.; Zhao, Y.; Zhou, R.; Zhang, S., Eltrombopag is a potential target for drug intervention in SARS-CoV-2 spike protein. *Infect. Genet. Evol.* **2020,** *85*, 104419.

36. Ruan, Z.; Liu, C.; Guo, Y.; He, Z.; Huang, X.; Jia, X.; Yang, T., SARS-CoV-2 and SARS-CoV: Virtual screening of potential inhibitors targeting RNA-dependent RNA polymerase activity (NSP12). *J Med Virol* **2020**.

37. Alexpandi, R.; De Mesquita, J. F.; Karutha Pandian, S. K.; Ravi, A. V., Quinolines-Based SARS-CoV-2 3CLpro and RdRp Inhibitors and Spike-RBD-ACE2 Inhibitor for Drug-Repurposing Against COVID-19: An in silico Analysis. *Front. Microbiol.* **2020**.

38. Durdagi, S.; Aksoydan, B.; Dogan, B.; Sahin, K.; Shahraki, A. Screening of Clinically Approved and Investigation Drugs as Potential Inhibitors of COVID-19 Main Protease: A Virtual Drug Repurposing Study *ChemRxiv* [Online], 2020. chemrxiv.12032712.v2.

39. Senathilake, K.; Samarakoon, S.; Tennekoon, K. Virtual Screening of Inhibitors Against Spike Glycoprotein of SARS-CoV-2: A Drug Repurposing Approach *Preprints* [Online], 2020. preprints202003.0042.v2.

40. Subramanian, S. Some FDA Approved Drugs Exhibit Binding Affinity as High as -16.0 Kcal/mol Against COVID-19 Main Protease (mpro): A Molecular Docking Study *IndiaRxiv* [Online], 2020. osf.io/t7jsd.

41. Sharma, K.; Morla, S.; Goyal, A.; Kumar, S., Computational guided drug repurposing for targeting 2'-O-ribose methyltransferase of SARS-CoV-2. *Life Sci* **2020**, 118169.

42. Peterson, L., In Silico Molecular Dynamics Docking of Drugs to the Inhibitory Active Site of SARS-CoV-2 Protease and Their Predicted Toxicology and ADME. *Available at SSRN 3580951* **2020**.

43. Adem, S.; Eyupoglu, V.; Sarfraz, I.; Rasul, A.; Ali, M. Identification of potent COVID-19 main protease (Mpro) inhibitors from natural polyphenols: An in silico strategy unveils a hope against CORONA *Preprints* [Online], 2020. preprints202003.0333.v1.

44. Chakraborti, S.; Bheemireddy, S.; Srinivasan, N. Repurposing drugs against main protease of SARS-CoV-2: mechanism based insights supported by available laboratory and clinical data 2020. chemrxiv.12057846.v2.

45. Wang, J., Fast identification of possible drug treatment of coronavirus disease-19 (COVID-19) through computational drug repurposing study. *J. Chem. Inf. Mod.* **2020,** *60* (6), 3277-3286.

46. Kouznetsova, V.; Huang, D.; Tsigelny, I. F., Potential COVID-19 Protease Inhibitors: Repurposing FDAapproved Drugs. **2020**.

47. Chen, Y. W.; Yiu, C.-P. B.; Wong, K.-Y., Prediction of the SARS-CoV-2 (2019-nCoV) 3C-like protease (3CL pro) structure: virtual screening reveals velpatasvir, ledipasvir, and other drug repurposing candidates. *F1000Research* **2020,** *9*, 129.

48. Arul, M. N.; Kumar, S.; Jeyakanthan, J.; Srivastav, V. Searching for target-specific and multi-targeting organics for Covid-19 in the Drugbank database with a double scoring approach *ResearchSquare* [Online], 2020. rs.3.rs-36233/v1.

49. Romeo, A.; Iacovelli, F.; Falconi, M., Targeting the SARS-CoV-2 spike glycoprotein prefusion conformation: virtual screening and molecular dynamics simulations applied to the identification of potential fusion inhibitors. *Virus Res* **2020,** *286*, 198068.

50. Xie, X.; Muruato, A. E.; Zhang, X.; Lokugamage, K. G.; Fontes-Garfias, C. R.; Zou, J.; Liu, J.; Ren, P.; Balakrishnan, M.; Cihlar, T.; Tseng, C.-T. K.; Makino, S.; Menachery, V. D.; Bilello, J. P.; Shi, P.-Y. A nanoluciferase SARS-CoV-2 for rapid neutralization testing and screening of anti-infective drugs for COVID-19 *bioRxiv* [Online], 2020. 2020.06.22.165712.

51. Sobeh, M.; Mrid, R. B.; Yasri, A. Virtual Screening of Heterocyclic Molecules to Identify Potential SARS-COV2 virus Mpro Protease Inhibitors for Further Medicinal Chemistry design *Res. Square* [Online]. rs.3.rs-37557/v1.

52. de Lima Menezes, G.; da Silva, R. A., Identification of potential drugs against SARS-CoV-2 non-structural protein 1 (nsp1). *J Biomol Struct Dyn* **2020**, 1-11.

53. Huang, D.; Yu, H.; Wang, T.; Yang, H.; Yao, R.; Liang, Z., Efficacy and safety of umifenovir for coronavirus disease 2019 (COVID-19): A systematic review and meta-analysis. *J Med Virol* **2020**.

54. Huang, H.; Guan, L.; Yang, Y.; Le Grange, J. M.; Tang, G.; Xu, Y.; Yuan, J.; Lin, C.; Xue, M.; Zhang, X.; Chen, R.; Zhou, L.; Huang, W. Chloroquine, arbidol (umifenovir) or lopinavir/ritonavir as the antiviral monotherapy for COVID-19 patients: a retrospective cohort study *ResearchSquare* [Online], 2020. rs.3.rs-24667/v1.

55. Naveen, S. M.; Reddy, M. S. Target SARS-CoV-2: Computation of Binding energies with drugs of Dexamethasone/Umifenovir by Molecular Dynamics using OPLS-AA force field *Res. Square* [Online], 2020. rs-40785/v1.

56. Alméciga-Díaz, C. J.; Pimentel-Vera, L. N.; Caro, A.; Mosquera, A.; Moreno, C. A. C.; Rojas, J. P. M.; Díaz-Tribaldos, D. C. Virtual screening of potential inhibitors for SARS-CoV-2 main protease *Preprints* [Online], 2020. preprints202004.0146.v1.

57. Khater, S.; Dasgupta, N.; Das, G. Combining SARS-cov-2 Proofreading Exonuclease and RNA-dependent RNA Polymerase Inhibitors as a Strategy to Combat COVID-19: A High-throughput in Silico Screen *OSF Preprints* [Online], 2020. osf.io/7x5ek.

58. Chakraborti, S.; Srinivasan, N. Drug Repurposing Approach Targeted Against Main Protease of SARS-CoV-2 Exploiting 'Neighbourhood Behaviour'in 3D Protein Structural Space and 2D Chemical Space of Small Molecules *ChemRxiv* [Online], 2020. chemrxiv.12057846.v1.

59. Fintelman-Rodrigues, N.; Sacramento, C. Q.; Ribeiro Lima, C.; Souza da Silva, F.; Ferreira, A. C.; Mattos, M.; de Freitas, C. S.; Cardoso Soares, V.; da Silva Gomes Dias, S.; Temerozo, J. R.; Miranda, M. D.; Matos, A. R.; Bozza, F. A.; Carels, N.; Alves, C. R.; Siqueira, M. M.; Bozza, P. T.; Souza, T. M. L., Atazanavir, alone or in combination with ritonavir, inhibits SARS-CoV-2 replication and pro-inflammatory cytokine production. *Antimicrob Agents Chemother* **2020**.

60. Borgio, J. F.; Alsuwat, H. S.; Al Otaibi, W. M.; Ibrahim, A. M.; Almandil, N. B.; Al Asoom, L. I.; Salahuddin, M.; Kamaraj, B.; AbdulAzeez, S., State-of-the-art tools unveil potent drug targets amongst clinically approved drugs to inhibit helicase in SARS-CoV-2. *Arch Med Sci* **2020,** *16* (3), 508-518.

61. Sekhar, T. Virtual Screening based prediction of potential drugs for COVID-19 *Preprints* [Online], 2020. preprints202002.0418.v2.

62. Omotuyi, O.; Nash, O.; Ajiboye, B.; Metibemu, D.; Oyinloye, B.; Ojo, A. The disruption of SARS-CoV-2 RBD/ACE-2 complex by Ubrogepant Is mediated by interface hydration *Preprints* [Online], 2020. preprints202003.0466.v1). .

63. Kandwal, S.; Fayne, D. Repurposing drugs for treatment of SARS-CoV-2 infection: Computational design insights into mechanisms of action *ResearchSquare* [Online], 2020. rs.3.rs-54535/v1.

64. Bembenek, S. Drug Repurposing and New Therapeutic Strategies for SARS-CoV-2 Disease Using a Novel Molecular Modeling-AI Hybrid Workflow *ChemRxiv* [Online], 2020. chemrxiv.12449081.v2.

65. Jiménez-Alberto, A.; Ribas-Aparicio, R. M.; Aparicio-Ozores, G.; Castelán-Vega, J. A., Virtual screening of approved drugs as potential SARS-CoV-2 main protease inhibitors. *Comp. Biol. Chem.* **2020**, 107325.

66. Yu, J.; Shao, S.; Liu, B.; Wang, Z.; Jiang, Y.-Z.; Li, Y.; Chen, F.; Liu, B. Emergency Antiviral Drug Discovery During a Pandemic-a Case Study on the Application of Natural Compounds to Treat COVID-19 *ChemRxiv* [Online], 2020. chemrxiv.12307592.v1.

67. Olubiyi, O. O.; Olagunju, M.; Keutmann, M.; Loschwitz, J.; Strodel, B., High throughput virtual screening to discover inhibitors of the main protease of the coronavirus SARS-CoV-2. *Preprints* **2020,** *10*.

68. Contreras-Puentes, N.; Alvíz-Amador, A., Virtual Screening of Natural Metabolites and Antiviral Drugs with Potential Inhibitory Activity against 3CL-PRO and PL-PRO. *Biomed. Pharmacol. J.* **2020,** *13* (2).

69. Ramírez-Salinas, G. L.; Martínez-Archundia, M.; Correa-Basurto, J.; García-Machorro , J. Repositioning of ligands that target spike glycoprotein as potential drugs against SARS-CoV-2 *ResearchSquare* [Online], 2020. rs.3.rs-52025/v1.